\documentclass[pdflatex,sn-mathphys-num]{sn-jnl}% Math and Physical Sciences Numbered Reference Style
\usepackage{graphicx}%
\usepackage{multirow}%
\usepackage{amsmath,amssymb,amsfonts}%
\usepackage{amsthm}%
\usepackage[switch]{lineno}
\usepackage{mathrsfs}%
\usepackage[title]{appendix}%
\usepackage{xcolor}%
\usepackage{textcomp}%
\usepackage{manyfoot}%
\usepackage{booktabs}%
\usepackage{algorithm}%
\usepackage{hyperref}
\usepackage{algorithmicx}%
\usepackage{algpseudocode}%
\usepackage{listings}%

\usepackage{subcaption}
\usepackage{tikz}
\usetikzlibrary{decorations.pathreplacing, arrows.meta, positioning, calc, shapes.geometric}
\definecolor{orcidlogocol}{HTML}{A6CE39}
\renewcommand{\orcidlogo}{%
  \tikz[baseline=-0.5ex]{\node[fill=orcidlogocol,circle,inner sep=0.5pt,text=white,font=\sffamily\bfseries\tiny] {iD};}%
}

\begin{document}
% \linenumbers

\title{Flow-Based Surrogates for High-Dimensional Likelihoods in Experimental Neutrino Physics}

\author*[1]{\fnm{Mathias} \sur{El Baz}\,\orcid{https://orcid.org/0000-0002-9562-3897}}\email{mathias.elbaz@unige.ch}
\equalcont{These authors contributed equally to this work.}

\author*[1]{\fnm{Lorenzo} \sur{Giannessi}\,\orcid{https://orcid.org/0000-0001-6899-7116}}\email{lorenzo.giannessi@unige.ch}
\equalcont{These authors contributed equally to this work.}

\author[2]{\fnm{Adrien} \sur{Blanchet}\,\orcid{https://orcid.org/0000-0002-4992-0161}}\email{adrien.blanchet@lpnhe.in2p3.fr}

\author[1]{\fnm{Federico} \sur{Sánchez}\,\orcid{https://orcid.org/0000-0003-0320-3623}}

\affil*[1]{\orgdiv{Section de Physique, DPNC}, \orgname{University of Geneva}, \orgaddress{\city{Geneva}, \country{Switzerland}}}

\affil[2]{\orgdiv{Laboratoire de Physique Nucléaire et de Hautes Énergies, LPNHE}, \orgname{CNRS/IN2P3, Sorbonne Université, Université Paris Cité}, \orgaddress{\city{Paris}, \country{France}}}


\maketitle

\begin{abstract}

Precision long-baseline neutrino experiments use near-detector data to constrain systematic uncertainties on the unoscillated neutrino flux, a prerequisite for accurate oscillation-parameter measurements at the far detector. When the constrained likelihood is high-dimensional and non-Gaussian, this procedure demands advanced statistical treatment. Here we show that normalizing flows provide faithful and portable likelihood models for this problem. Leveraging an initial Gaussian approximation of the likelihood, we train a hybrid architecture combining coupling transformations and autoregressive spline flows. We demonstrate the method on a representative near-detector likelihood replica with 110 systematic uncertainty parameters, 10 of which explicitly introduce non-Gaussianities in the constrained likelihood. The trained model achieves a relative effective sample size of $98\%$, compared with about $5\%$ for the Gaussian approximation, and reproduces a Markov Chain Monte Carlo reference while remaining closed-form, samplable, and pointwise evaluable, making it suitable to uncertainty propagation in downstream physics analyses.

\end{abstract}

\section{Introduction}\label{sec1}

Long-baseline neutrino-oscillation experiments aim to measure the leptonic mixing angles, the neutrino mass ordering, and the charge–parity (CP) violating phase that may help explain the matter–antimatter asymmetry of the Universe~\cite{t2k_cp_2020}. As these measurements sharpen, the experiments have entered a precision regime in which the limiting factor is often no longer statistical power alone, but the faithful modeling and propagation of systematic uncertainties. Flux predictions, neutrino-nucleus interaction models, detector efficiency and reconstruction effects are constrained through likelihood fits, in which nuisance parameters are adjusted so that the predicted event rates best describe control data. These fits can involve large numbers of nuisance parameters, whose impact on predicted spectra is often correlated and non-Gaussian~\cite{T2K:2011qtm,T2K:2013bqz,ankowski_2017,dune_tdr_physics_2020,t2k_oscillation_2023,dilodovico_2023,t2k_mixing_2025,t2k_nd_systematics_2026}. Moreover, these fits depart from the familiar setting in which a few parameters of interest are isolated by marginalizing away the rest. Here both the parameters of interest and the nuisance parameters are numerous, so eliminating the nuisances is a genuinely hard problem~\cite{eliminating_nuisance}, and the constrained uncertainties it yields must enable or enhance downstream correlated analyses, as happens in the near-detector--far-detector fit scheme typical of long-baseline experiments.
In this context, the quantity that must be carried from one stage of the analysis to the next is the constrained likelihood over the systematic parameters.

Within a single long-baseline experiment, near-detector constraints must be propagated to the far-detector prediction before the oscillation parameters are extracted. More demanding still, recent combinations across experiments, such as the joint T2K-NOvA oscillation analysis and the Super-Kamiokande-T2K atmospheric-accelerator fit, require the constraints obtained in one analysis to enter the likelihood of another~\cite{t2k_sk_joint_2025,t2k_nova_joint_2025,nufit2024,dune_configs_2021}. In such scenarios, a fit result is only as useful as the fidelity with which its constrained likelihood can be represented and reused. This mirrors a broader effort in particle physics to publish full statistical models rather than summary results~\cite{Cranmer_2022}, so that the information content of a fit is preserved.

The question addressed here is therefore how this constrained likelihood should be exported for downstream tasks. By a portable likelihood representation, we mean a compact object that can be distributed independently of the original analysis software, evaluated at arbitrary points in parameter space, sampled efficiently, and inserted into downstream likelihoods without rerunning the upstream fit or accessing the original data. Such a representation must preserve the relevant non-Gaussian structure of the original likelihood while remaining practical to use in far-detector predictions, joint fits, or global analyses.

In current oscillation analyses, this propagation is often performed by replacing the constrained likelihood with a local Gaussian approximation around its maximum~\cite{t2k_oscillation_2023,t2k_mixing_2025,t2k_nd_systematics_2026}. This so-called post-fit Gaussian is centered on the best-fit nuisance parameters, with a covariance matrix inferred from the curvature of the likelihood at that point.
 This representation is compact and easy to reuse, offering a continuous closed-form likelihood, but it assumes that the log-likelihood is quadratic and that the relevant systematic variations are adequately described by an ellipsoid in parameter space. Non-linear parameter dependencies, physical boundaries, limited-statistics effects~\cite{BARLOW1993219}, and degeneracies can instead generate asymmetric marginals, curved correlations, and non-Gaussian tails. Markov Chain Monte Carlo methods~\cite{10.1063/1.1699114, 10.1093/biomet/57.1.97} provide a more faithful sampling of such likelihoods, but a finite chain is not a closed-form continuous likelihood function, therefore limiting the practicality of its use for constraint propagation. Full experiment-specific likelihood implementations provide the most faithful representation, but require access to dedicated software, detailed analysis input, and experiment-specific infrastructure, making them difficult to distribute or insert into independent downstream analyses~\cite{Cranmer:1456844}. This leaves a methodological gap between Gaussian summaries, which are portable but potentially biased, MCMC samples, which are faithful but not directly evaluable, and full likelihood implementations, which are accurate but heavy.

In this work, we address this gap using normalizing flows~\cite{Kobyzev_2021, JMLR:v22:19-1028} (NF) as explicit surrogate models for high-dimensional systematic likelihoods. Normalizing flows transform a simple base distribution through a sequence of invertible learnable maps, producing a density model that is both directly samplable and pointwise evaluable. Once trained, the flow is not only a generator of nuisance-parameter samples, but a continuous surrogate for the constrained likelihood itself: it can be sampled, evaluated, and inserted into downstream likelihoods on which gradient-based optimization can be performed, without rerunning the original fit nor accessing the original data. Flow-based and, more broadly, simulation-based inference methods have become a widely adopted tool across the physical sciences~\cite{doi:10.1073/pnas.1912789117, brehmer_cranmer_2020, brehmer_2021, tejero_cantero_2020}. Applications include posterior estimation in gravitational-wave astronomy and cosmology~\cite{green_2020, dax_2021, dax_2023}, fast and high-fidelity detector simulation~\cite{krause_caloflow_2021, krause_caloflow2_2021}, and asymptotically exact sampling of lattice field theories~\cite{albergo_2019, kanwar_2020, albergo_2021}. Within neutrino physics, related ideas have been explored for oscillation inference, event generation and detector-calibration likelihood modeling~\cite{pina_otey_2020, PhysRevD.102.013003, PhysRevD.109.032008, gavrikov_2026, icecube_flow_2026, v7sz-vn3b}. 

We demonstrate this approach on a realistic high-dimensional statistical-analysis replica inspired by the near-detector fit of the currently running T2K experiment, where the modeling of systematic uncertainties involves $\mathcal{O}(10^2)$ fit parameters constrained by near-detector data. This benchmark provides a controlled setting in which the target likelihood is known, non-Gaussian features can be isolated, and direct comparisons can be made with both the post-fit Gaussian approximation and MCMC sampling. We study the accuracy of the trained flow as a likelihood surrogate, its sampling efficiency, and its impact on propagated flux predictions. Finally, we discuss its possible use as a portable constrained likelihood model for downstream oscillation analyses.

\section{Results}

\subsection{Normalizing flows as likelihood surrogates}
\label{sec:nf}

Our aim is to find a continuous density model of the likelihood function that can be sampled, evaluated, and used for downstream tasks. The target density is the normalized systematic likelihood,
\begin{equation}
p(\boldsymbol{\eta}) \propto \mathcal{L}(\boldsymbol{\eta}),
\end{equation}
where $\boldsymbol{\eta}$ denotes the vector of systematic uncertainty parameters. The normalizing-flow surrogate~\cite{Kobyzev_2021} learns an invertible map $f_\theta:\mathbb{R}^D\to\mathbb{R}^D$ that transports a simple base distribution to this target density. In practice, it is convenient to use the post-fit Gaussian approximation as a starting point. It captures the local structure of the likelihood around the best-fit region and offers an efficient way to generate an initial training dataset. The post-fit Gaussian therefore plays a dual role: it serves both as an initial proposal distribution and as a source of training samples. However, the method does not intrinsically rely on training samples drawn from the Gaussian approximation. The samples used to initialize the training can be chosen according to the expected structure of the constrained likelihood, and may instead be obtained from other sources, such as an existing MCMC chain. In the present application, samples from the post-fit Gaussian provide a convenient starting point.

\begin{figure}[t]
\centering
\begin{tikzpicture}[
scale=0.85,
  >=Latex,
  font=\footnotesize,
  blk/.style   ={draw=black!75, line width=0.5pt, rounded corners=2pt,
                  align=center, inner sep=5pt,
                  minimum height=12mm, minimum width=34mm, fill=white},
  io/.style    ={draw=black!50, line width=0.4pt, rounded corners=2pt,
                  align=center, inner sep=4pt, fill=white},
  arr/.style   ={->, line width=0.55pt, >=Latex, draw=black!85},
  edge/.style  ={line width=0.55pt, draw=black!85},
  cond/.style  ={->, line width=0.55pt, >=Latex, draw=black!55, densely dashed},
  lbl/.style   ={font=\scriptsize, inner sep=1.5pt, fill=white},
  dot/.style   ={circle, fill=black!85, inner sep=0pt, minimum size=3pt},
]

\node[io]  (u)     at (0,0)
  {$u\sim p_u$\\[1pt]\scriptsize$\mathcal{N}(\mathbf{0},I)$};
\node[blk] (lin)   at (3.1,0)
  {Linear flow\\[2pt]$x = T^{(1)}_{\theta_1}(u)$};
\node[blk] (cl)    at (8.0, 1.5)
  {Coupling-layer flow\\[2pt]$\boldsymbol{\eta}_B = T^{(2)}_{\theta_2}(x_B)$};
\node[blk] (arblk) at (8.0,-1.5)
  {Autoregressive flow\\[2pt]
   $\boldsymbol{\eta}_A = T^{(3)}_{\theta_3}(x_A;\,\boldsymbol{\eta}_B)$};
\node[io]  (out)   at (13.5,0)
  {$(\boldsymbol{\eta}_A,\boldsymbol{\eta}_B)$\\[1pt]\scriptsize$\sim q_{\mathrm{NF}}$};

\draw[arr] (u) -- (lin);

\coordinate (sp) at ($(lin.east)+(0.18,0)$);
\draw[edge] (lin.east) -- (sp);
\node[dot] at (sp) {};
\draw[arr] (sp) -- (cl.west);
\draw[arr] (sp) -- (arblk.west);
\node[lbl, above] at ($(sp)!0.55!(cl.west)$)    {$x_B$};
\node[lbl, below] at ($(sp)!0.55!(arblk.west)$) {$x_A$};

\coordinate (mrg) at (12.0,0);
\draw[arr] (cl.east)    -| (mrg);
\draw[arr] (arblk.east) -| (mrg);
\node[dot] at (mrg) {};
\draw[arr] (mrg) -- (out);
\node[lbl, above] at ($(cl.east)   +(0.95,0)$) {$\boldsymbol{\eta}_B$};
\node[lbl, below] at ($(arblk.east)+(0.95,0)$) {$\boldsymbol{\eta}_A$};

\coordinate (tap)   at ($(cl.east)+(0.45,0)$);
\coordinate (cMidV) at ($(tap)+(0,-1.2)$);
\coordinate (cMidH) at (cMidV -| arblk.north);
\node[dot, scale=0.7] at (tap) {};
\draw[cond] (tap) -- (cMidV) -- (cMidH) -- (arblk.north);
\node[lbl, above] at ($(cMidV)!0.5!(cMidH)$) {cond.};

\end{tikzpicture}
\caption{Hybrid normalizing-flow architecture with level-of-detail approach used to model
$q_{\mathrm{NF}}(\boldsymbol{\eta}_A,\boldsymbol{\eta}_B)
=q_{\mathrm{AR}}(\boldsymbol{\eta}_A\!\mid\!\boldsymbol{\eta}_B)\,
q_{\mathrm{CL}}(\boldsymbol{\eta}_B)$.
Base samples $u\sim\mathcal{N}(\mathbf{0},I)$ pass through a linear flow
$T^{(1)}_{\theta_1}$, optionally initialized from the Gaussian post-fit
$(\hat{\boldsymbol{\eta}},C)$ via $\mu_{\theta_1}\!=\!\hat{\boldsymbol{\eta}}$
and $L_{\theta_1}L_{\theta_1}^{\!\top}\!=\!C$. Its output is partitioned as
$x=(x_A,x_B)$. The high-dimensional easy block is mapped to
$\boldsymbol{\eta}_B$ by a coupling-layer flow $T^{(2)}_{\theta_2}$; the
low-dimensional difficult block is mapped to $\boldsymbol{\eta}_A$ by a
conditional autoregressive flow $T^{(3)}_{\theta_3}$, with
$\boldsymbol{\eta}_B$ feeding the autoregressive conditioner networks
(dashed arrow).}
\label{fig:architecture_schematic}
\end{figure}
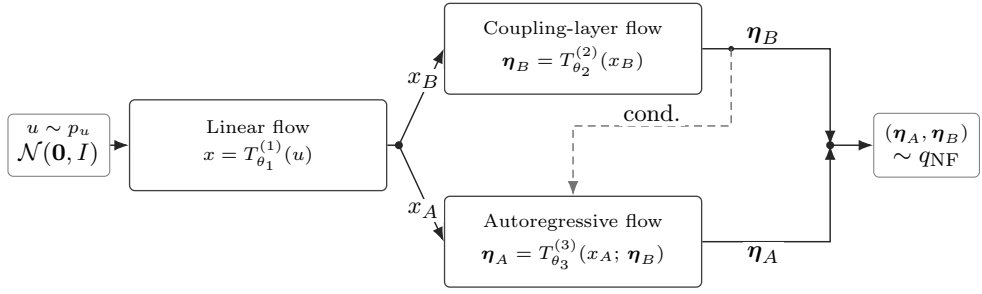

The surrogate is designed to exploit the structure of the systematic-parameter likelihood. The large, approximately linear sector of the parameter space is modeled with coupling layers~\cite{dinh2017densityestimationusingreal}, while the smaller sector containing the most non-linear systematic parameters is modeled with a conditional autoregressive spline flow~\cite{papamakarios2017masked,winkler2023learninglikelihoodsconditionalnormalizing}. This keeps the expensive autoregressive structure restricted to the directions where non-Gaussian effects are expected to be strongest, while preserving correlations with the remaining high-dimensional block. The corresponding factorization is
\begin{equation}
q_{\rm NF}(\boldsymbol{\eta}_A,\boldsymbol{\eta}_B)
=
q_{\rm AR}(\boldsymbol{\eta}_A\mid\boldsymbol{\eta}_B)
q_{\rm CL}(\boldsymbol{\eta}_B),
\label{eq:results_hybrid_factorization}
\end{equation}
where $\boldsymbol{\eta}_A$ denotes the non-linear block and $\boldsymbol{\eta}_B$ the high-dimensional approximately linear block. This architecture is summarized in Figure~\ref{fig:architecture_schematic}, and the full implementation is given in the Methods section~\ref{nf_model}.

\subsection{Replicating a typical near-detector likelihood}
\label{sec:replica}

To benchmark the method, we construct a realistic near-detector likelihood~\cite{t2k_nd_systematics_2026} replica designed to reproduce the statistical structure of a typical long-baseline neutrino near-detector analysis. The purpose of this benchmark is to build a controlled likelihood with similar dimensionality and complexity: a large number of correlated systematic parameters, non-linear event-weight variations, finite-statistics fluctuations, and a post-fit covariance matrix obtained from a maximum-likelihood fit.

The simulated samples consist of neutrino and antineutrino interactions on hydrocarbon, generated with a neutrino-energy spectrum inspired by the T2K off-axis beam. The selected samples mimic the topology-based categories used in near-detector analyses, with three samples in forward horn current (FHC) neutrino mode and three samples in reverse horn current (RHC) antineutrino mode, classified according to the number of detected final-state pions. A statistically independent simulated sample is used as pseudo-data. The pseudo-data and the Monte Carlo model are generated using the same underlying neutrino-nucleus interaction model. Details of the event generation, detector smearing, selection thresholds, sample composition, and likelihood construction are given in the Methods section~\ref{app:demo_dataset}.

The likelihood is parametrized with $110$ parameters representing a set of hypothetical systematic uncertainties. Of these, $100$ act as flux-like systematics: they rescale the predicted event rates through binned (50 bins for each beam mode) weights in true neutrino energy, with the event-rate variation in any given analysis bin being linear in the parameters. The remaining $10$ are non-linear parameters that emulate interaction-model uncertainties via event-by-event weights depending on truth-level kinematic variables. Because these weights distort the model histograms in ways that do not need to be linear in $\boldsymbol{\eta}$, they can produce likelihood profiles that depart from Gaussianity. Details on the parametrization of the systematic uncertainties model are provided in the Methods section~\ref{app:syst_param}.

The scale is large enough to reproduce the non-Gaussian structure of a realistic near-detector likelihood and the separation between a high-dimensional approximately linear bulk and a small non-linear block, while remaining a setup for which a self-contained replica can be built from scratch and a well-converged MCMC reference can be obtained for a controlled comparison. The relevant computational scaling of the NF model is governed primarily by the size of the non-linear block, whose impact is quantified in Section~\ref{sec:sampling_efficiency}.

\subsection{Fidelity in parameter space}
\label{sec:validation}

We first verify that the trained flow reproduces the target likelihood in the parameter space. As a faithful reference we use a Markov Chain Monte Carlo (MCMC) sampling of the same likelihood, generated with a Metropolis-Hastings algorithm~\cite{10.1063/1.1699114, 10.1093/biomet/57.1.97}. Being asymptotically exact, the MCMC provides the reference against which the flow is validated throughout this work. The chain consists of $4.2\times10^{6}$ samples, of which the first $2\times10^{5}$ are discarded as burn-in. The proposal was tuned during burn-in to target the theoretical optimal acceptance probability of $0.234$ for random-walk Metropolis--Hastings~\cite{roberts1997weak} (the corresponding optimum is $0.651$ for Hamiltonian Monte Carlo~\cite{beskos2013optimal}).

\begin{figure}
    \centering
    \includegraphics[width=0.99\linewidth]{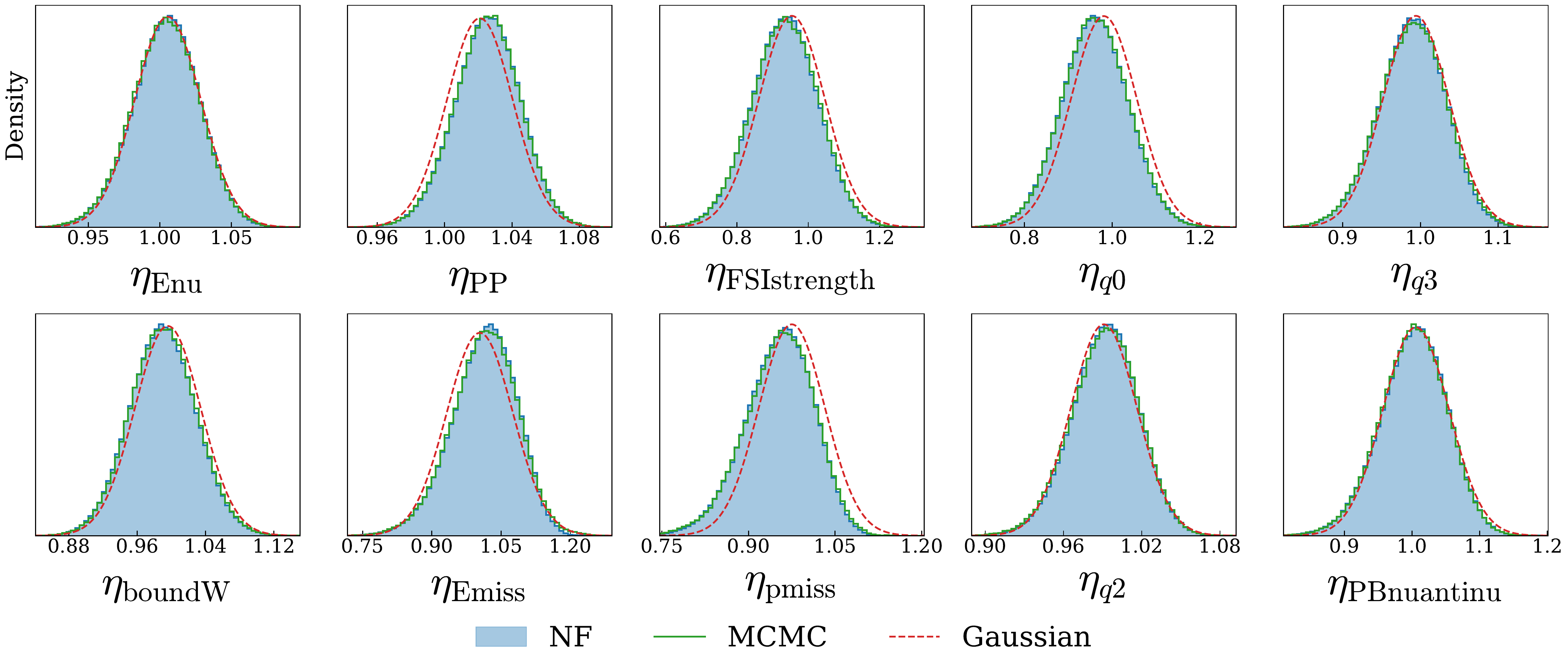}
    \caption{One-dimensional marginal distributions of the ten non-linear systematic parameters, obtained from $500\,000$ samples drawn from the trained NF surrogate, $500\,000$ samples from the MCMC reference, and the post-fit Gaussian approximation. See Methods~\ref{app:syst_param} for the parameter definitions.}
    \label{fig:marginal}
\end{figure}

Figure~\ref{fig:marginal} compares the one-dimensional marginal distributions of the ten non-linear systematic parameters, obtained from $500\,000$ samples drawn from the trained NF surrogate, $500\,000$ samples from the MCMC reference, and the analytical post-fit Gaussian approximation. These parameters are the ones expected to develop the strongest departures from Gaussianity. The NF marginals closely follow the MCMC reference across all ten parameters, reproducing their asymmetric tails and modes displaced from the best-fit value, while the Gaussian approximation, symmetric and centered on the best-fit point by construction, fails to capture them.

\begin{figure}[h!]
  \centering
  \includegraphics[width=0.79\textwidth]{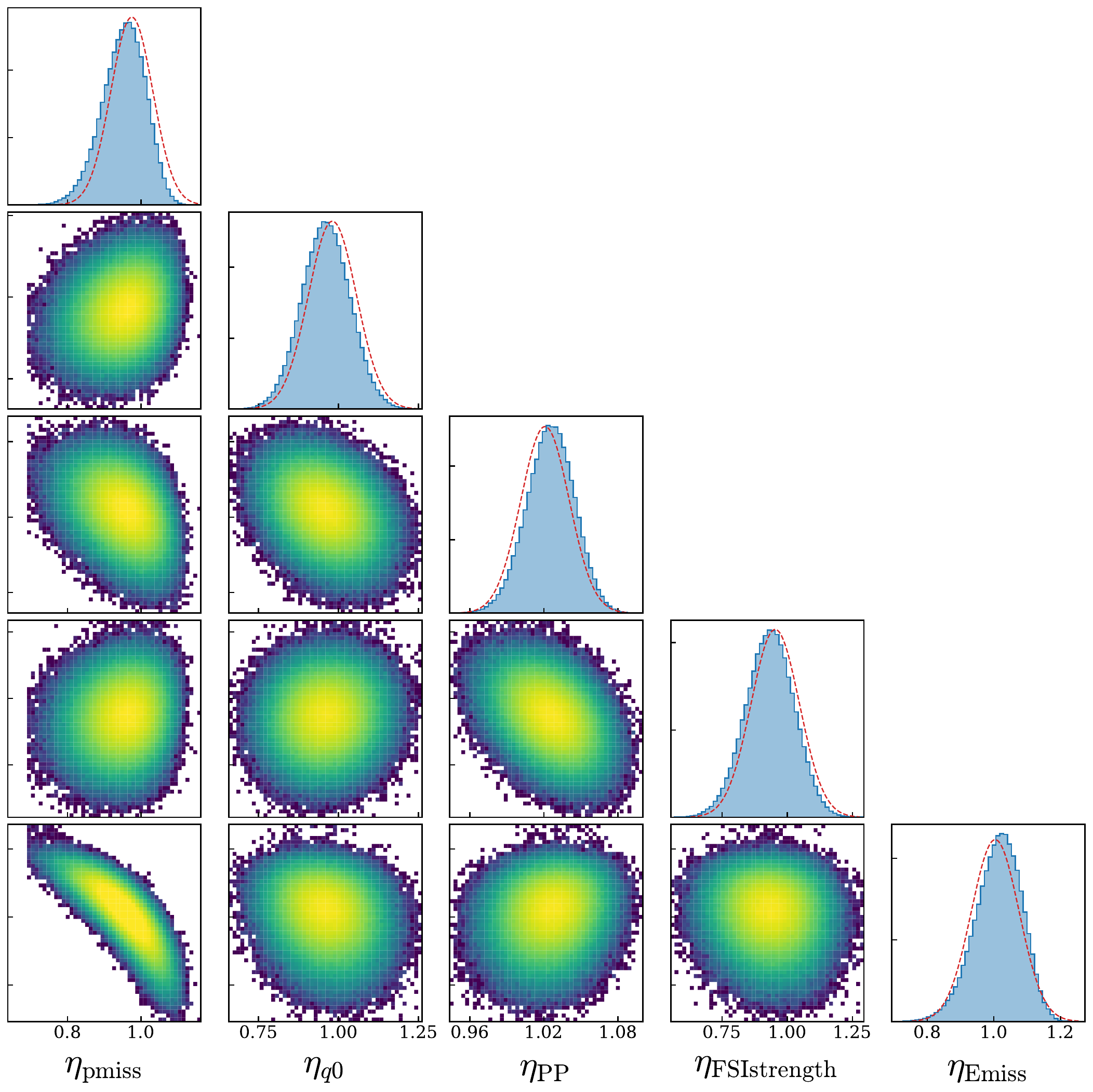}
    \caption{Corner plot of selected non-linear systematic parameters obtained from $500\,000$ samples drawn from the NF surrogate. The diagonal panels show the one-dimensional marginal distributions, compared with the post-fit Gaussian approximation shown as a red dashed curve. The off-diagonal panels show the corresponding two-dimensional projections in logarithmic color scale. See Method~\ref{app:syst_param} for the parametrization of the nuisance parameters.}
  \label{fig:corner_plot}
\end{figure}

The non-Gaussian features are most visible in two dimensions: Figure~\ref{fig:corner_plot}, from $500\,000$ NF samples, shows curved correlations and distorted confidence regions in the off-diagonal panels, for instance between the missing-momentum and missing-energy parameters.

Moreover, the maximum of each marginal distribution of the NF model is shifted with respect to the Gaussian and in agreement with the MCMC. This highlights the potential of the NF model in reducing shortcomings of the Gaussian approximation in terms of the description of the bulk probability mass, reducing inference bias.

\subsection{Flux predictions}

The trained normalizing flow can be used as a directly samplable surrogate of the full systematic-parameter likelihood, allowing flux predictions to be produced and compared with those obtained from a MCMC sampling of the same likelihood. For each sampled vector of systematic parameters $\boldsymbol{\eta}$, the corresponding systematic response is propagated to the event weights. The reweighted events are then binned in true neutrino energy, separately for samples of neutrino-beam (FHC) and antineutrino-beam (RHC), producing one flux prediction for each systematic throw. Repeating this procedure for samples drawn from the NF surrogate, the post-fit Gaussian approximation, and the MCMC chain gives an ensemble of flux predictions from which the mean, uncertainty bands, and bin-to-bin correlations can be estimated.

\begin{figure}[h!]
  \centering
  \begin{minipage}{0.49\textwidth}
    \centering
    \includegraphics[width=\textwidth]{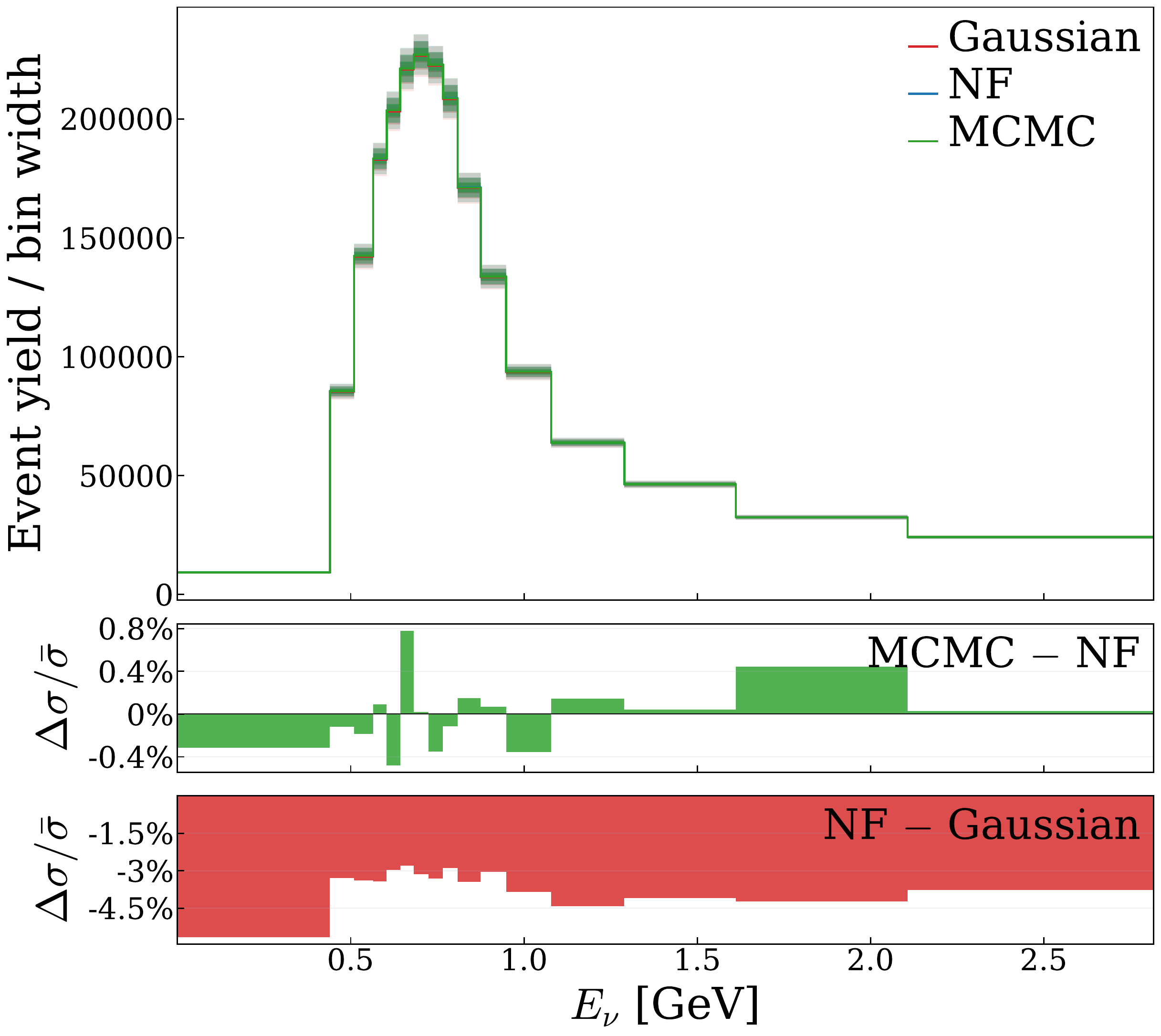}
  \end{minipage}
  \hfill
  \begin{minipage}{0.49\textwidth}
    \centering
    \includegraphics[width=\textwidth]{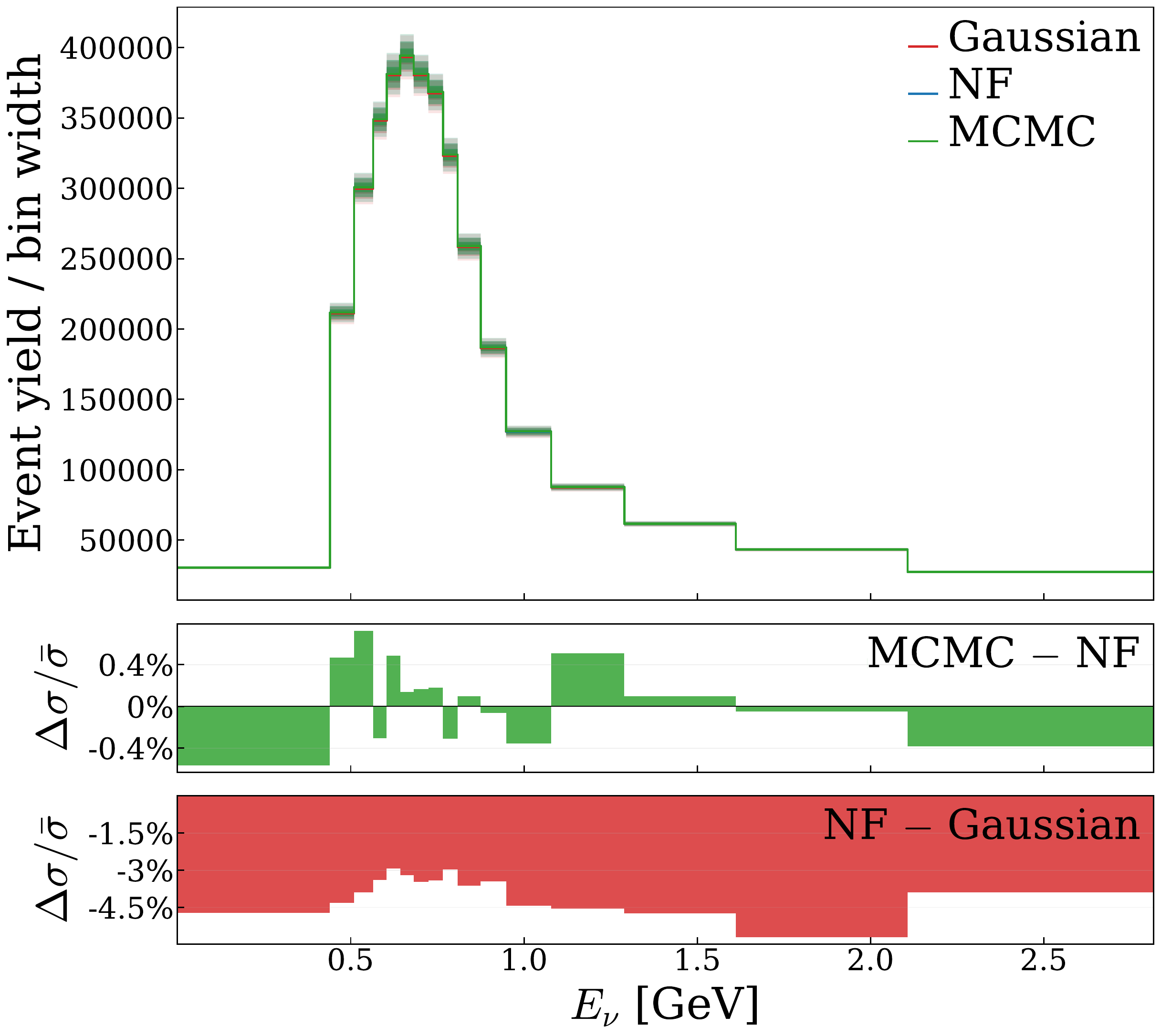}
  \end{minipage}

  \caption{FHC (left) and RHC (right) flux predictions obtained by propagating systematic-parameter throws to the event weights and filling the resulting $E_\nu$ spectra. The predictions are shown as the event yield divided by the bin width, using $500\,000$ samples drawn from the NF surrogate, $500\,000$ MCMC steps, and the corresponding post-fit Gaussian mean and standard deviation. The spectra are truncated at $E_\nu=2.82~\mathrm{GeV}$, and the binning is chosen such that the expected nominal population is approximately equal in each bin. The lower panels compare the relative uncertainty widths, $\Delta\sigma/\bar{\sigma}$, between MCMC and NF and between NF and the Gaussian approximation, where $\bar{\sigma}$ is the average of the two uncertainties being compared.}
  \label{fig:flux_spectrum_sigma}
\end{figure}

Figure~\ref{fig:flux_spectrum_sigma} shows the unoscillated flux predictions obtained with each likelihood model (NF, MCMC and Gaussian) for the FHC and RHC samples. The upper panels show the mean event yield divided by the bin width, together with bin-by-bin uncertainty bands estimated as the sample standard deviation of the bin event yield over a repetition of toy experiments. The lower panels compare the relative uncertainty widths, defined as
\[
\frac{\Delta \sigma}{\bar{\sigma}}
=
\frac{\sigma_A-\sigma_B}{(\sigma_A+\sigma_B)/2},
\]
where $\sigma_A$ and $\sigma_B$ are the bin-by-bin standard deviations obtained from the flux predictions of two different methods. The NF and MCMC uncertainty estimates agree within approximately $\pm 1\%$, while the post-fit Gaussian approximation overestimates the uncertainty width by about $3.5\%$ compared to both NF and MCMC. This indicates that the NF reproduces the uncertainty propagation obtained from the reference MCMC sample. This has direct consequences for the inference of the parameters of interest: the NF-based propagation provides a more accurate estimate of the systematic uncertainty, and in this particular case leads to a smaller propagated uncertainty than the post-fit Gaussian approximation. 

\begin{figure}[h!]
  \centering
  \includegraphics[width=0.95\textwidth]{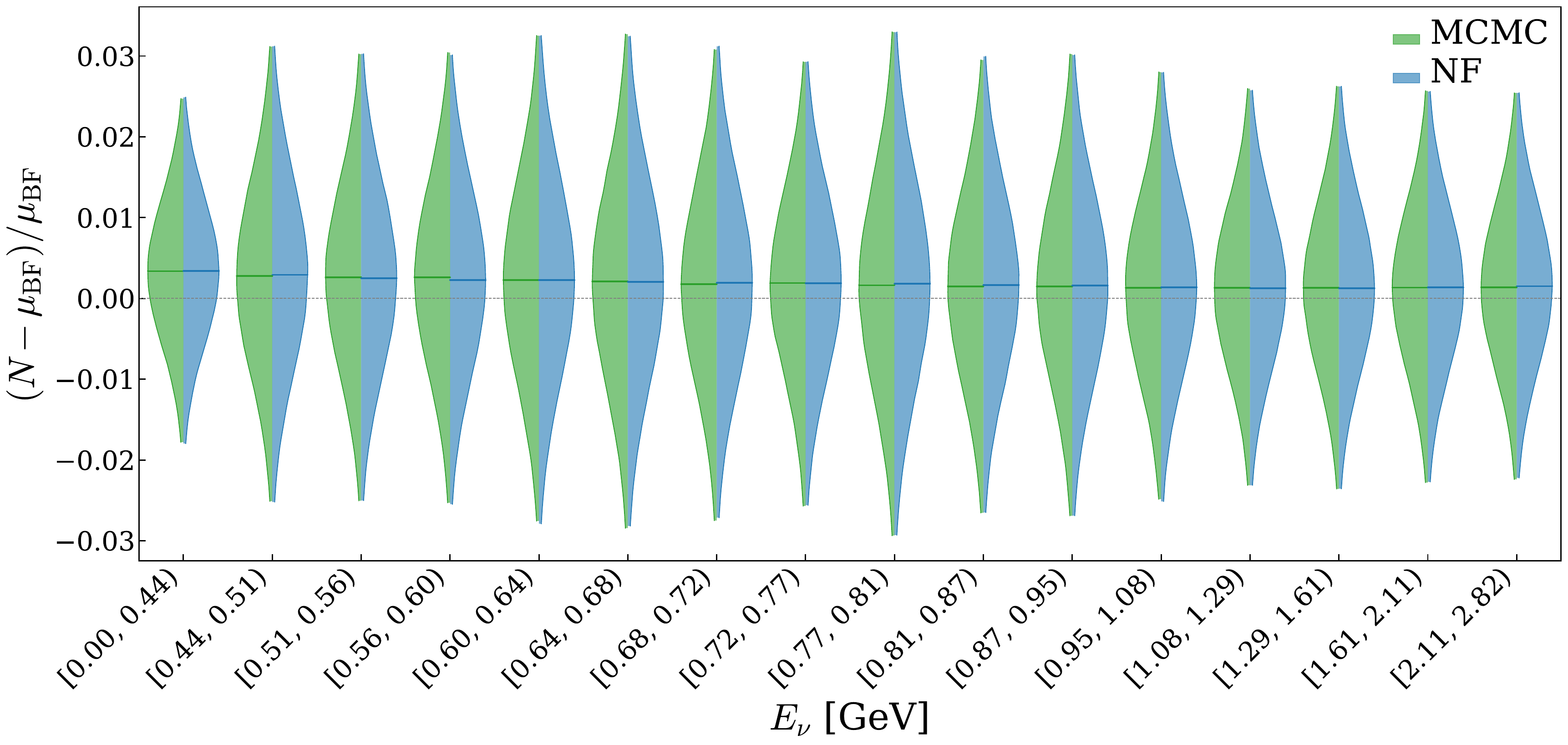}
    \caption{Split violin representation of the FHC flux predictions in the same $E_\nu$ bins, using $500\,000$ samples drawn from the NF surrogate and $500\,000$ samples from the MCMC chain. For each systematic-parameter throw, the event yield $N$ in a given bin is compared with the fixed post-fit Gaussian prediction $\mu_{\mathrm{BF}}$ in that bin. Each violin shows the resulting distribution of the relative deviation $(N-\mu_{\mathrm{BF}})/\mu_{\mathrm{BF}}$ over throws, with the MCMC distribution shown on the left side and the NF distribution on the right side of the violin. The horizontal segment drawn inside each half-violin indicates the median of the corresponding distribution, while the dashed horizontal line at zero marks the post-fit Gaussian reference.}
    \label{fig:flux_violin}
\end{figure}

 Figure~\ref{fig:flux_violin} provides a complementary bin-by-bin comparison using split violin plots. Each violin shows the relative deviation of the bin event yields with respect to the Gaussian mean prediction. The two distributions are in very good agreement in each bin: they exhibit the same shift, a similar spread, and a similar distribution shape. In particular, both the MCMC and NF distributions exhibit a median shifted by approximately $+0.2\%$ relative to the Gaussian reference, indicating an underestimation of the median event rate by the Gaussian approximation. This shows that using the post-fit Gaussian as a surrogate for the systematic likelihood could introduce a bias in the predictive estimate of the flux after systematic uncertainty propagation, while the NF reproduces the MCMC results at the bin-by-bin distribution level.

\subsection{Sampling efficiency and computational cost}
\label{sec:sampling_efficiency}

\begin{figure}
    \centering
    \includegraphics[width=0.7\linewidth]{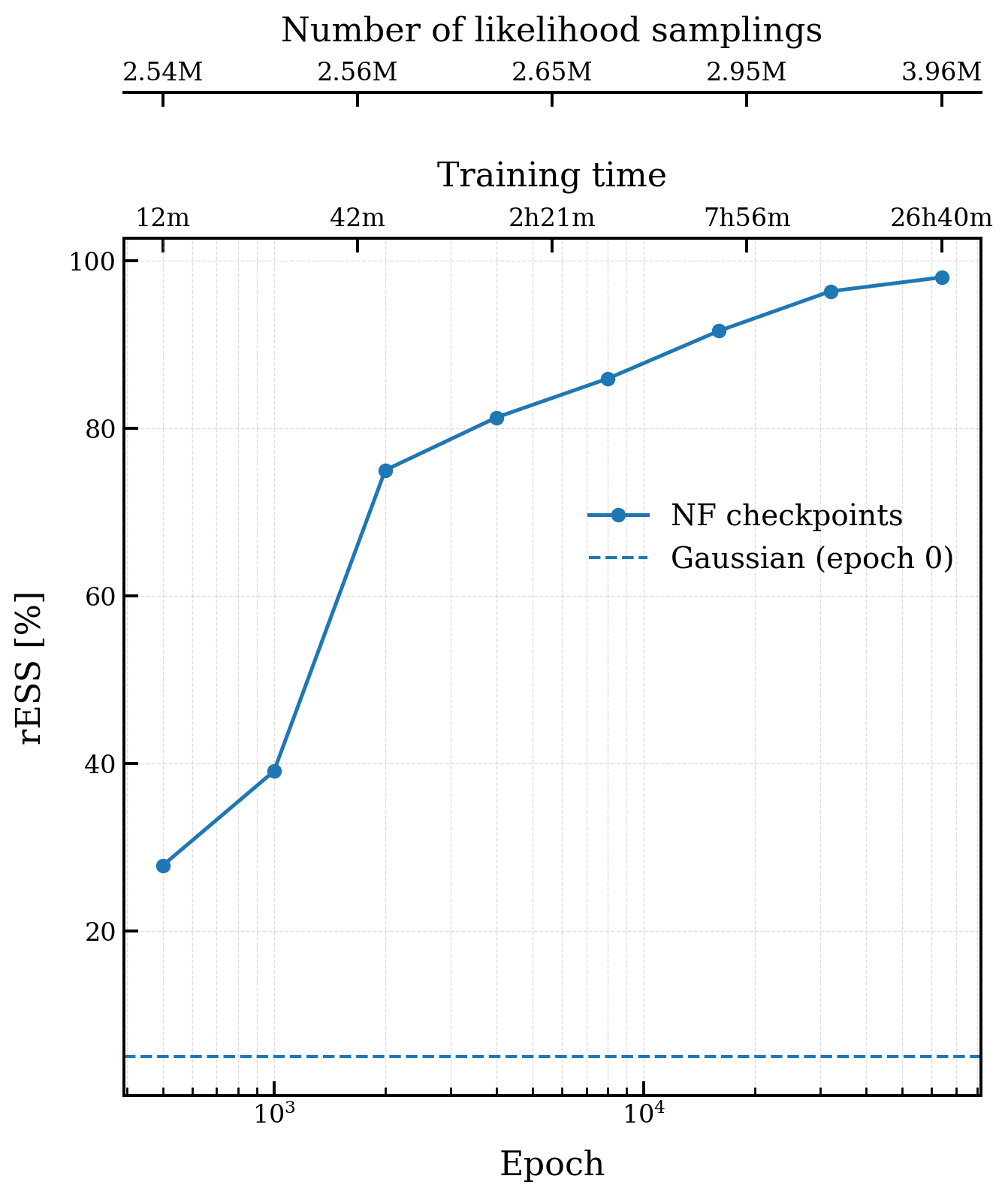}
    \caption{Relative effective sample size of a set of 500\,000 samples from the current NF model at different times during training.}
    \label{fig:ESS}
\end{figure}

Figure~\ref{fig:ESS} shows the evolution of the sampling quality during training, quantified by the relative effective sample size,
\[ \mathrm{rESS} = \frac{1}{N} \frac{\left(\sum_{i=1}^{N} w_i\right)^2}{\sum_{i=1}^{N} w_i^2}, \]
where $N$ is the number of samples and $w_i$ is the importance weight associated with the $i$-th sample. For samples drawn from a proposal distribution $q$, the weights are defined as
\[ w_i = \frac{p(\boldsymbol{\eta}_i)}{q(\boldsymbol{\eta}_i)}, \]
where $p$ denotes the target likelihood distribution. During training, $q$ corresponds to the distribution represented by the current normalizing flow, while the Gaussian baseline is obtained by taking $q=g(\boldsymbol{\eta};\boldsymbol{\eta}_{\mathrm{best\text{-}fit}},C)$ where $\boldsymbol{\eta}_{\mathrm{best\text{-}fit}}$ and $C$ are the post-fit maximum likelihood estimator and covariance matrix.

\begin{figure}[h!]
  \centering
  \includegraphics[width=0.79\textwidth]{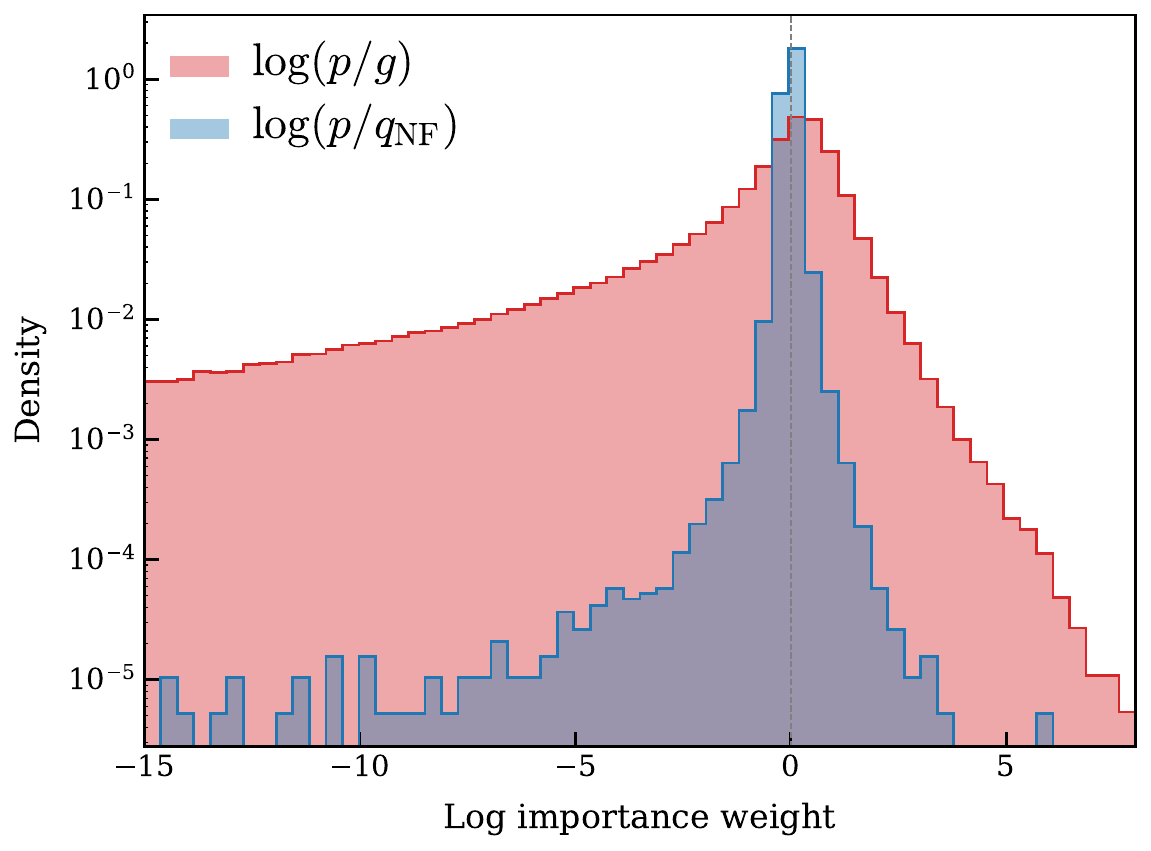}
  \caption{Distribution of the log-importance weights for $500\,000$ samples drawn from the post-fit Gaussian proposal $g$ (red) and from the trained normalizing flow $q_{\rm NF}$ (blue). }
  \label{fig:weight_histogram}
\end{figure}

The rESS obtained with the Gaussian approximation is shown as a dashed line for comparison. When the training is stopped after $64\,000$ epochs, the NF reaches a rESS of $98\%$, compared to only $5\%$ for the Gaussian proposal. This corresponds to an overall improvement of approximately a factor $20$ in sampling efficiency when replacing the initial post-fit Gaussian proposal with the trained normalizing flow. The corresponding distribution of importance weights is shown in Figure~\ref{fig:weight_histogram}, obtained using $500\,000$ samples drawn from the trained NF and $500\,000$ samples drawn from the post-fit Gaussian approximation. While the Gaussian proposal produces weights spanning more than $10$ orders of magnitude, the NF weights are much more concentrated. This shows that the trained NF provides a much closer approximation to the target likelihood distribution, leading to substantially more efficient importance sampling. 

Moreover, once trained, the surrogate becomes essentially independent of the expensive likelihood evaluation. For the benchmark configuration with $D_A=10$, both sampling and density evaluation are analytical and fast, reaching approximately $38$ million samples and density evaluations per GPU-hour, or $960\,000$ per CPU-hour. This should be compared with the original likelihood, for which only about $10\,000$ evaluations per CPU-hour are possible on the same hardware.

As anticipated in Section~\ref{sec:nf}, the sampling cost of the trained flow is mainly controlled by the autoregressive block. It therefore scales with $D_A$, the number of non-linear parameters modeled autoregressively, rather than with the total dimension. By contrast, the coupling block is evaluated in a single pass. The relevant quantity for a given analysis is therefore the number of genuinely non-linear systematics, which is typically a small fraction of the total: the T2K 2023 oscillation analysis, for instance, used $59$ non-linear parameters out of 711~\cite{t2k_oscillation_2023}. Table~\ref{tab:split_scaling} reports the sampling throughput as $D_A$ is
increased up to this scale. In the worst case scenario $D_A=100$, the trained surrogate generates of order $10^4$ samples per CPU-hour and $10^6$ per GPU-hour, still exceeding the $\sim\!10^4$ evaluations per CPU-hour of the original likelihood. 

The computational cost of training the NF model is dominated by the evaluation of the likelihood, which also drives the computational cost of likelihood sampling methods like MCMC. Figure~\ref{fig:ESS} highlights that the number of likelihood evaluations required to obtain an efficient NF model is comparable to the number of points sampled by the benchmark MCMC. Unlike MCMC, the NF samples are not autocorrelated, and the training cost is paid only once: the trained flow is a closed-form density that can be sampled and evaluated thereafter at marginal cost.

\begin{table}[t]
\centering
\begin{tabular}{ccc}
\hline
$D_A$ & CPU [samples/hr] & GPU [samples/hr] \\
\hline
$10$  & $9.6\times10^{5}$ & $3.8\times10^{7}$ \\
$20$  & $2.3\times10^{5}$ & $1.4\times10^{7}$ \\
$30$  & $1.2\times10^{5}$ & $7.3\times10^{6}$ \\
$60$  & $5.6\times10^{4}$ & $2.0\times10^{6}$ \\
$100$ & $2.2\times10^{4}$ & $8.4\times10^{5}$ \\
\hline
\end{tabular}
\caption{Sampling speed of the trained flow as a function of the dimension $D_A$ of the autoregressive block, at fixed total dimension.}
\label{tab:split_scaling}
\end{table}

\section{Discussion}

We have shown that a normalizing flow can accurately reproduce the
high-dimensional constrained systematic likelihood of a realistic
T2K-inspired near-detector benchmark with $110$ parameters, including its
asymmetric and non-quadratic features. Once trained, the normalizing flow turns the fitted systematic likelihood into a portable object: it can be sampled, evaluated, and propagated without rerunning the original fit. As in the case of a post-fit Gaussian approximation, the original data and full fitting machinery are no longer needed for downstream propagation, but the NF preserves the non-Gaussian structure of the likelihood with much higher accuracy. The important step is a change of representation: the fit output is promoted to an explicit density model that can be efficiently evaluated and sampled wherever it is needed.

This places the NF in a useful middle ground between MCMC samples and Gaussian likelihood surrogates. Compared with MCMC, it is lightweight, explicit, and directly samplable: new systematic throws can be generated on demand, without storing large chains or recomputing the original likelihood. Compared with a Gaussian approximation, it retains skewness, curved correlations, asymmetric marginals, and non-Gaussian tails, and therefore avoids the bias that can appear when a likelihood is propagated through an overly restrictive quadratic model. 

Such a representation is particularly useful when the constrained systematic likelihood from one analysis must be propagated into another, as is the case for oscillation-parameter measurements using a two-step near-detector/far-detector approach, as well as joint experiment fits, or global combinations involving several detectors, experiments, or data samples, fundamental for precision oscillation measurements in the near future~\cite{t2k_oscillation_2023,t2k_nova_joint_2025,nufit2024}. In such scenarios, an NF surrogate can be inserted as an explicit, systematics density model: it can be sampled for uncertainty propagation, used for nuisance marginalization, or evaluated as a prior or constraint term in a downstream likelihood on which gradient descent can be performed, all without relying on a Gaussian approximation of the original fit. As these analyses grow in scale and ambition, the ability to carry a likelihood in its full complexity will be no less essential than the precision of the fit that produced it.

\section{Methods}

\subsection{Benchmark near-detector likelihood}
\label{app:demo_dataset}

\begin{figure}
    \centering
    \includegraphics[width=0.49\linewidth]{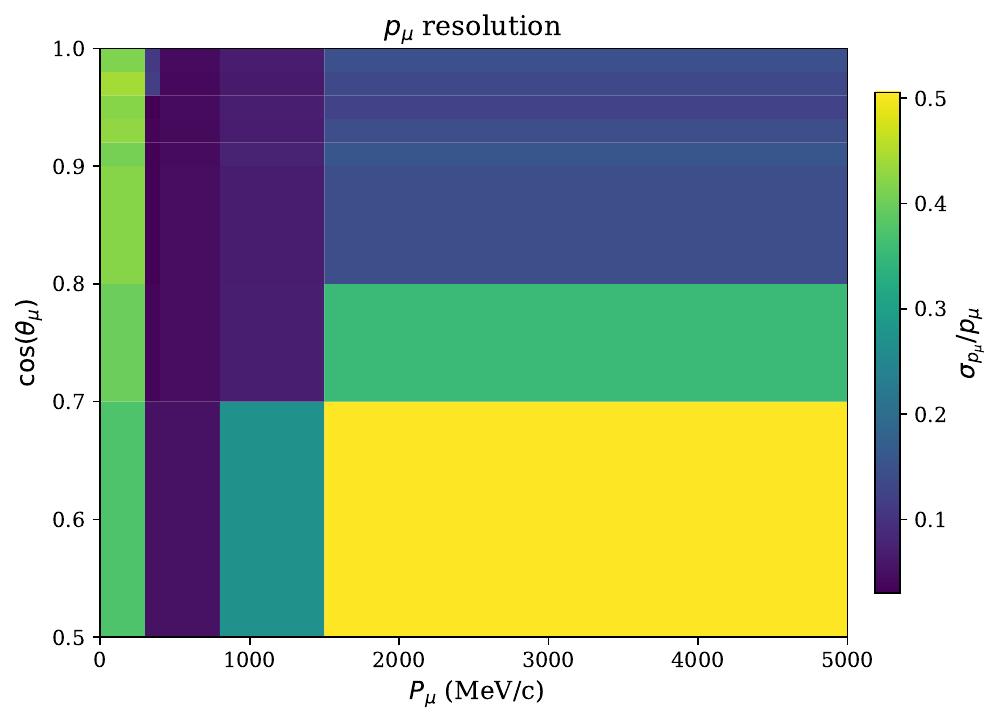}
    \includegraphics[width=0.49\linewidth]{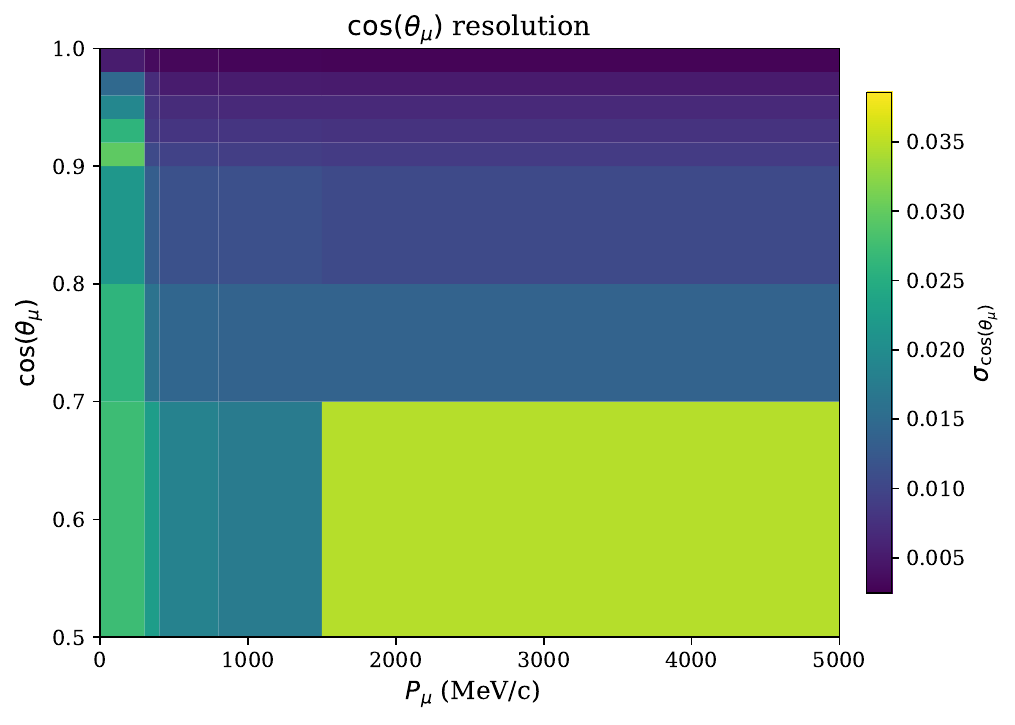}
    \caption{Simulated muon-kinematics detector resolution as a function of $p_\mu$ and $\cos\theta_\mu$. Left: muon momentum resolution. Right: muon angular resolution.}
    \label{fig:sigma_pmucosthetamu}
\end{figure}

We benchmark the method by simulating an end-to-end near-detector-like analysis. We generate 500\,000 neutrino and 500\,000 antineutrino interactions on hydrocarbon using the NEUT~\cite{NEUT2021} framework and formatted as NUISANCE~\cite{nuisance} flat trees~\cite{Dolan:2026nlr}. Since the benchmark is inspired by T2K near-detector analyses, the simulation follows the most recent publicly available T2K off-axis neutrino-energy spectrum~\cite{T2K_FLUX}, peaking at approximately 0.6~GeV. Detector response is simulated heuristically by applying Gaussian smearing to the final-state muon kinematics of each simulated event. We use resolutions on the muon momentum magnitude $|\vec p_\mu|$ and on the cosine of the muon scattering angle $\cos\theta_\mu$ similar to those of the T2K Near Detector (ND280)~\cite{T2K:2011qtm, ND280_TPC, ND280_FGD}, as shown in Figure~\ref{fig:sigma_pmucosthetamu}.

Typical near-detector data samples are produced using topological event selections, where each event is assigned to a sample based on the number of detected particles in the final state. In order to simulate detector-efficiency effects, we introduce fixed thresholds for the detection of final-state charged particles, summarized in Table~\ref{tab:thresholds}.

\begin{table}
    \centering
    \begin{tabular}{ccc}
    \hline
        Particle & Momentum & $\cos(\theta)$\\
        \hline
        Muon & 10~MeV & 0.5 \\
        Proton & 450~MeV & 0.5\\
        Pion & 200~MeV & 0.5\\
        \hline
    \end{tabular}
    \caption{Particle detection thresholds.}
    \label{tab:thresholds}
\end{table}

Six samples are defined: three samples in neutrino mode, FHC for Forward Horn Current, and three equivalent samples in antineutrino mode, RHC for Reverse Horn Current. The samples are defined according to the number of detected charged pions in the final state: no pion, one pion, or more than one pion. Exactly one detected muon (antimuon) is required in each neutrino (antineutrino) sample to pass the selection. The total efficiency of the selection is around 55\%. Table~\ref{tab:samples} summarizes the number of selected events in each sample. The binning is optimized for each sample based on expected event rates, with higher granularity in high-statistics regions.

\begin{table}[h]
\centering
\begin{tabular}{lr}
\hline
\textbf{Sample} & \textbf{\# of events} \\
\hline
FHC $\nu_{\mu}$ CC $0\pi$          & 188\,305 \\
FHC $\nu_{\mu}$ CC $1\pi$          &  43\,154 \\
FHC $\nu_{\mu}$ CC Other           &   4\,272 \\
RHC $\bar{\nu}_{\mu}$ CC $0\pi$    & 280\,545 \\
RHC $\bar{\nu}_{\mu}$ CC $1\pi$    &  27\,666 \\
RHC $\bar{\nu}_{\mu}$ CC Other     &   1\,100 \\
\hline
\textbf{Total}                     & 545\,042 \\
\hline
\end{tabular}
\caption{Summary of selected events per Monte Carlo sample.}
\label{tab:samples}
\end{table}

To emulate the procedure of fitting a Monte Carlo model to data, we repeat the same generation and selection procedure on an independent sample containing the same number of neutrino and antineutrino interactions, which we label as pseudo-data.

\subsection{Likelihood construction}
\label{app:likelihood_construction}

The likelihood is written as a function of the systematic-parameter vector $\boldsymbol{\eta}$,
\begin{equation}
    -2\ln\mathcal{L}(\boldsymbol{\eta})
    =
    -2\ln\mathcal{L}_{\rm stat}(\boldsymbol{\eta})
    +
    \chi^2_{\rm syst}(\boldsymbol{\eta}),
    \label{eq:LH}
\end{equation}
where $\mathcal{L}_{\rm stat}$ compares the observed bin counts to the predicted event rates, and $\chi^2_{\rm syst}$ encodes the prior constraints on the systematic parameters. With prior mean $\boldsymbol{\eta}_0$ and prior covariance matrix $V$, the systematic penalty is
\begin{equation}
    \chi^2_{\rm syst}(\boldsymbol{\eta})
    =
    (\boldsymbol{\eta}-\boldsymbol{\eta}_0)^{\top}
    V^{-1}
    (\boldsymbol{\eta}-\boldsymbol{\eta}_0).
    \label{eq:syst_lh}
\end{equation}
The statistical term is built from a binned likelihood including Monte Carlo statistical uncertainties through the ``lite'' Barlow--Beeston prescription~\cite{BARLOW1993219,Conway:2011in}. The likelihood is evaluated with the GUNDAM analysis framework~\cite{gundam}, a tool used for oscillation and cross-section analyses in the T2K, ICARUS and DUNE experiments, which computes it for a given systematic-parameter vector $\boldsymbol{\eta}$.

The statistical term compares observed counts $n_i$ to the predicted mean $\mu_i(\boldsymbol{\eta})$ in each sample bin. The total statistical contribution is the sum over all bins,
\begin{equation}
    -2\ln\mathcal{L}_{\rm stat}(\boldsymbol{\eta})
    =
    -2\sum_i\ln\mathcal{L}_{\rm BB}^{i}(\boldsymbol{\eta}),
\end{equation}
where $\mathcal{L}_{\rm BB}^{i}$ denotes the ``lite'' Barlow--Beeston likelihood in bin $i$, as described in Ref.~\cite{Conway:2011in}. 

The likelihood's shape depends both on the statistical term in Eq.~\eqref{eq:LH} and on the way in which the parameters $\boldsymbol{\eta}$ modify the event rates.

The $110$ parameters are associated with an equal number of random variables, whose prior distribution is described by a $110\times110$ covariance matrix $V$, which enters the systematic term of the likelihood (Eq.~\eqref{eq:syst_lh}).

\subsection{Systematic-uncertainty parametrization}
\label{app:syst_param}

The systematic uncertainties are implemented as a set of parameters $\boldsymbol{\eta}$ that modify the underlying model histograms as a function of physically relevant quantities. The model is modified through event-by-event weights that depend on the value of the systematic parameters and on the true-level kinematics of the event, denoted generically as
\[
w(\boldsymbol{\eta},\mathrm{kinematics}).
\]
This provides model control through physics-motivated systematic parameters.

In a realistic analysis, such weights can be computed from variations of the cross-section model with respect to the baseline prediction through cross-section ratios,
\[
w(\eta_i)
=
\dfrac{
\sigma_{\text{var}}(\eta_i,\mathrm{kinematics})
}{
\sigma_{\text{baseline}}(\mathrm{kinematics})
}.
\]
The function $w(\eta_i)$ is obtained by evaluating the cross-section ratio at different values of $\eta_i$ and interpolating the result, for instance with polynomial splines. In the present benchmark, this behavior is mimicked with custom continuous functions of the systematic parameter and of the event kinematics.

The shape of the likelihood function $\mathcal{L}(\boldsymbol{\eta})$ is affected by the shape of the total event weight,
\[
w(\boldsymbol{\eta})=\prod_i w(\eta_i).
\]
Even in the limit of large statistics, the likelihood function is not necessarily Gaussian in $\boldsymbol{\eta}$ when some parameters enter the prediction through non-linear weight functions.

We distinguish between linear and non-linear systematic parameters. Linear parameters apply bin-by-bin multiplicative corrections to the predicted event distributions. For a parameter value $\eta_i$, the event yield in the corresponding bin is scaled by $w_i=\eta_i$. Non-linear weight functions mimic interaction-model uncertainties and are implemented as
\begin{equation}
    w
    =
    1+\alpha\, t(k)\left(\eta^p-\eta^p_{\text{prior}}\right),
    \label{eq:spline_onesided}
\end{equation}
where $\eta$ is the value of the systematic parameter. For simplicity, the dependence on event kinematics is taken to be a linear function of one kinematic variable $k$,
\begin{equation}
t(k)
=
2\left(
\frac{k-k_{\text{min}}}{k_{\text{max}}-k_{\text{min}}}
-0.5
\right).
\label{eq:t_k}
\end{equation}
The parameters $\alpha$, $p$, and the kinematic range $[k_{\text{min}},k_{\text{max}}]$ determine how the predicted spectra, and consequently the likelihood, vary as a function of the systematic parameter. They are defined separately for each non-linear parameter. The physically motivated parameters have imposed boundaries, which determine the domain of the likelihood function.

The framework is completed by a positive definite covariance matrix that encodes prior parameter uncertainties and correlations, together with a prior parameter vector $\boldsymbol{\eta}_{\text{prior}}$. This is typical of near-detector analyses, where linear parameters, such as detector and flux systematics, induce fluctuations of the spectra as a function of binned observables and can display correlations between adjacent bins, while correlations between cross-section uncertainties are motivated by the interaction model. No prior correlation is assumed between the two sets, although correlations may arise in the post-fit covariance matrix due to the way in which both sets affect the sample spectra~\cite{PhysRevD.109.072006}.

Each non-linear systematic parameter assigns a weight $w$ to each event based on the event kinematics and on the value of the systematic parameter $\eta$ as described in equations~\eqref{eq:spline_onesided} and~\eqref{eq:t_k}. In a realistic case, the event weight may depend on several kinematic variables, while in this benchmark each parameter depends on a single kinematic variable. Table~\ref{tab:splines} summarizes the parametrization of the non-linear systematic uncertainties. All prior values are set to 1.0 for simplicity, without loss of generality.

\begin{table}
    \centering
    \begin{tabular}{l p{5cm} l}
        \hline
        Parameter name & Kinematic dependence & $\alpha$ \\
        \hline
        \texttt{Enu} & Neutrino energy $E_\nu$  & 0.3 \\
        \texttt{PP} & Total energy of final-state pions & 0.3 \\
        \texttt{q0} & Energy transfer $q^0$ & 0.1 \\
        \texttt{q3} & Magnitude of momentum transfer $|\vec q|$ & 0.3 \\
        \texttt{FSIstrength} & Strength of final-state interactions, measured as the amount of energy lost by the proton before escaping the nucleus & 0.3 \\
        \texttt{W} & Hadronic invariant mass $W$  & 0.3 \\
        \texttt{Emiss} & Nuclear missing energy $E_{\text{miss}}$ & 0.1 \\
        \texttt{pmiss} & Nuclear missing momentum $p_{\text{miss}}$ & 0.1 \\
        \texttt{Q2} & Four-momentum transfer squared $Q^2$ & 0.1 \\
        \texttt{PBnuantinu} & Event rate modulation at low outgoing proton momentum, with opposite effects on $\nu$ and $\overline\nu$.& 0.1 \\
        \hline
    \end{tabular}
    \caption{Characteristics of the non-linear systematic-parameter implementation.}
    \label{tab:splines}
\end{table}

\begin{figure}[htbp]
    \centering
    \begin{subfigure}[b]{0.49\textwidth}
        \centering
        \includegraphics[width=\textwidth]{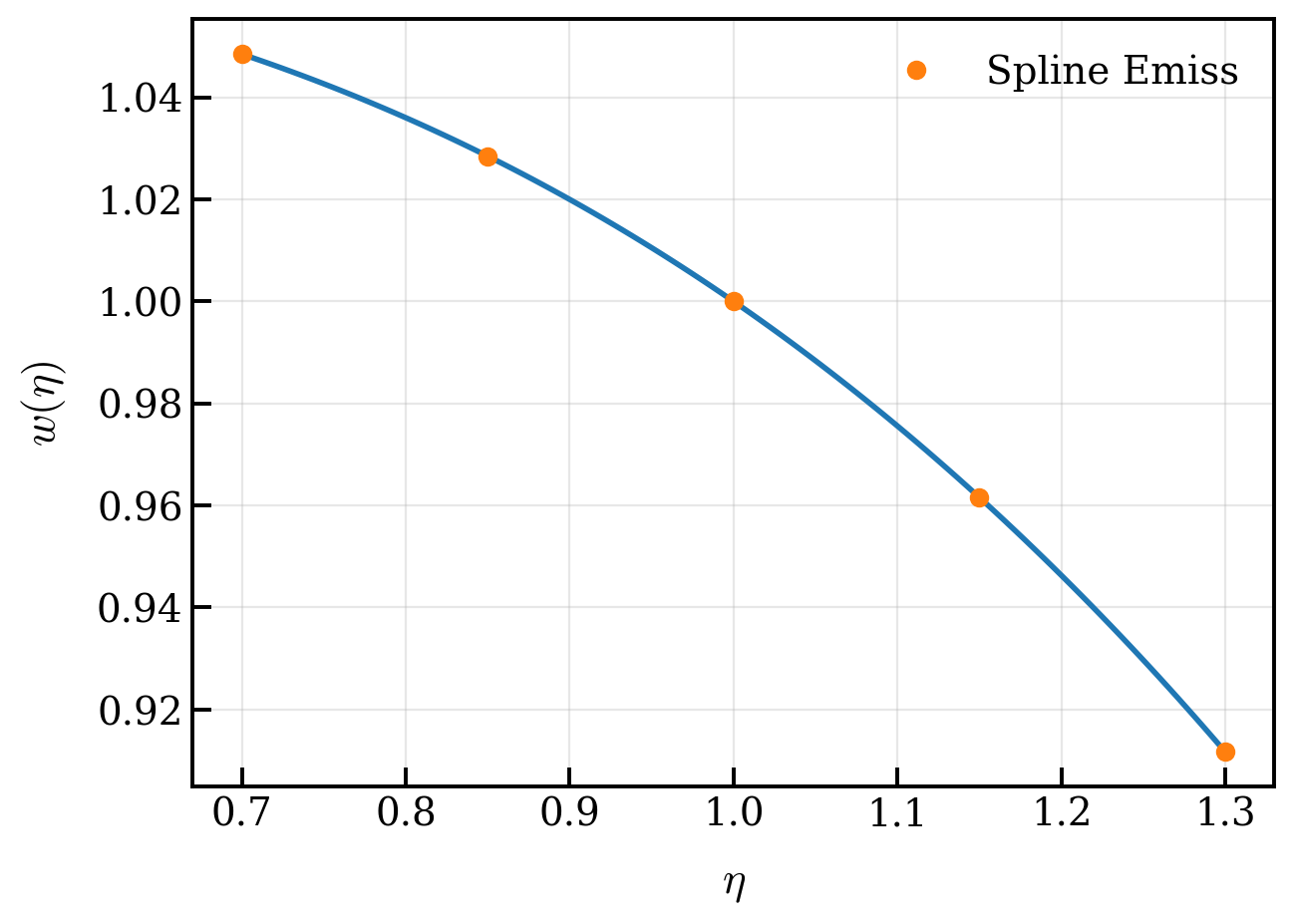}
        \caption{}
        \label{fig:sub1}
    \end{subfigure}
    \begin{subfigure}[b]{0.49\textwidth}
        \centering
        \includegraphics[width=\textwidth]{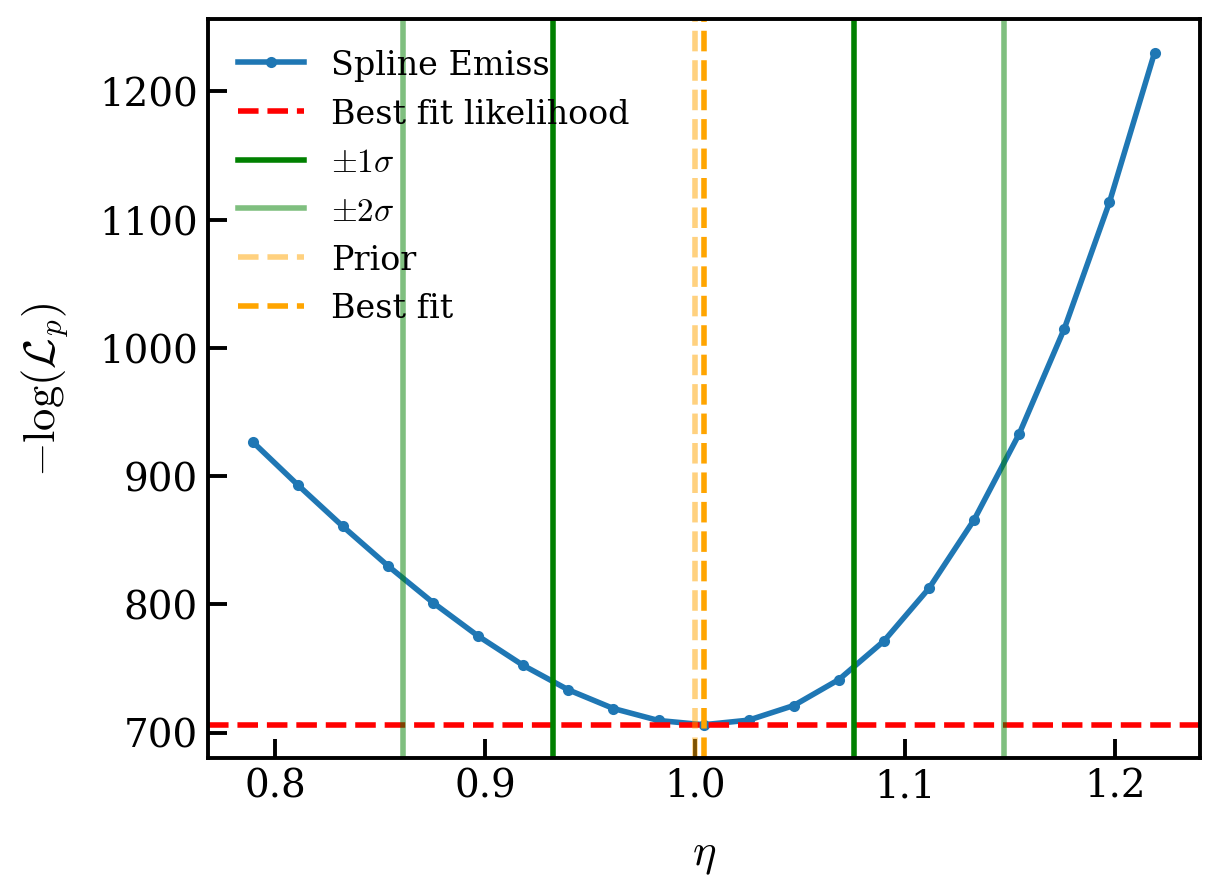}
        \caption{}
        \label{fig:sub2}
    \end{subfigure}
    
    \caption{(a): Event weight $w(\eta)$ as a function of the parameter value $\eta$. The points mark the values at which the weight is evaluated, interpolated by the continuous curve. \\ (b): Negative-log-likelihood profile obtained by scanning the $E_{\mathrm{miss}}$ parameter around its prior value, with all other parameters fixed. The dashed orange lines mark the prior and best-fit values, the green lines the $\pm1\sigma$ and $\pm2\sigma$ intervals, and the dashed red line the best-fit likelihood.}
    \label{fig:splines_and_scans}
\end{figure}

Figure~\ref{fig:splines_and_scans} shows how a representative non-linear weight function affects the likelihood through a negative-log-likelihood scan. For non-linear parameters, the NLL profiles can deviate from the quadratic shape expected for a Gaussian likelihood.

\subsection{Normalizing-flow model}\label{nf_model}

\subsubsection{Autoregressive flows and coupling layers}

Normalizing flows rely on invertible transformations that allow for efficient application of the change-of-variable formula. Two types of architectures show the best performance: autoregressive flows and coupling layers. Both approaches exploit triangular (or block-triangular) Jacobians to ensure computational efficiency, with complexity scaling linearly with the dimensionality of the probability space.  However, the choice between autoregressive flows and coupling layer-based flows ultimately depends on the application. 

In practice, autoregressive flows such as masked autoregressive flow (MAF)~\cite{papamakarios2017masked} are $D$ times slower to invert than to evaluate (or vice-versa), where $D$ is the dimension of the probability space.   On the other hand, flows based on coupling layers, such as NICE~\cite{dinh2015nicenonlinearindependentcomponents} or RealNVP~\cite{dinh2017densityestimationusingreal},  have an analytic one-pass inverse.  Sampling and evaluating the density at the same time requires evaluating both the forward and inverse transformations. Therefore, coupling layers allow both fast sampling and fast density evaluation. However, coupling layer-based flows are generally less expressive than autoregressive flows. 

We also exclude continuous normalizing flows~\cite{grathwohl2018ffjordfreeformcontinuousdynamics,lipman2023flowmatchinggenerativemodeling}, including flow-matching variants: although they scale well and can be easier to train, evaluating their log density requires integrating an augmented ODE for the divergence of the learned vector field, which is prohibitively expensive given the many repeated density evaluations our application requires.

\subsubsection{Parametrization of the flow transformation}

In addition to choosing an architecture (autoregressive or coupling-layer based), a normalizing flow must specify the family of invertible elementary transformations used to map between the base variables and the target parameter space. In this work we employ two complementary parametrizations: (i) Rational Quadratic Neural Spline Flows (RQ-NSF)~\cite{durkan2019neural,delbourgo1983c} for the autoregressive (conditional) block, and (ii) affine coupling blocks for the coupling-layer block. This choice reflects a trade-off between expressiveness and computational cost: RQ-NSF provide highly flexible, non-linear monotone transforms well suited to capture non-Gaussian structure, while affine coupling layers provide very fast sampling and density evaluation in high dimensions~\cite{Coccaro_2024}.

\paragraph{Rational Quadratic Neural Spline Flows}

RQ-NSF~\cite{durkan2019neural} model each transformed coordinate with a strictly monotone, piecewise rational--quadratic spline on a bounded interval $[-B,B]$, acting as the identity outside it. A neural network outputs the bin widths, heights, and internal-knot derivatives; within each bin $\log|T'(x)|$ is available in closed form and the inverse $T^{-1}$ follows from solving a quadratic, so both density evaluation and inversion are analytic~\cite{delbourgo1983c}.

The $D_A$-dimensional ``non-linear'' block $\boldsymbol{\eta}_A$ is transformed one dimension at a time using such 1D spline transformers. Writing $u_A$ for the base variables and $\eta_A$ for the transformed variables, a conditional autoregressive RQ-NSF layer takes the form
\begin{equation}
\eta_{A,i} \;=\; T_i\!\Big(u_{A,i};\,\psi_i(\,u_{A,<i},\,\boldsymbol{\eta}_B\,)\Big),
\qquad i=1,\ldots,D_A,
\label{eq:ar_rqnsf}
\end{equation}
where the parameters $\psi_i$ are produced by a conditioner network that depends on previous components (autoregressive masking) and, in our hybrid setting, also on the conditioning variables $\boldsymbol{\eta}_B$. The Jacobian of~\eqref{eq:ar_rqnsf} is triangular, hence the log-determinant reduces to a sum of 1D terms,
\begin{equation}
\log \left|\det \frac{\partial \boldsymbol{\eta}_A}{\partial \mathbf{u}_A}\right|
\;=\;
\sum_{i=1}^{D_A} \log \left| \frac{\partial T_i(u_{A,i};\psi_i)}{\partial u_{A,i}} \right|.
\end{equation}
RQ-NSF are particularly attractive because they combine (i) strong expressiveness for non-Gaussian targets with (ii) stable monotonicity constraints and (iii) tractable density evaluation and inversion~\cite{durkan2019neural}. They have been successfully applied within T2K for related inference tasks, including cross-section modeling and low-dimensional likelihood estimation~\cite{PhysRevD.102.013003,PhysRevD.109.032008}.

\paragraph{Affine coupling blocks}
\label{subsec:affine_coupling}

For the easy, high-dimensional block $\boldsymbol{\eta}_B$, we use affine coupling layers~\cite{dinh2017densityestimationusingreal}, which provide an analytic one-pass inverse and a cheap log-determinant even in very large dimensions. Splitting the input $z\in\mathbb{R}^{D_B}$ into two disjoint subsets $B_1$ and $B_2$, an affine coupling block leaves $z_{B_1}$ unchanged and transforms
\begin{equation}
z'_{B_2} = z_{B_2} \odot \exp\!\big(s_\theta(z_{B_1})\big) + t_\theta(z_{B_1}),
\label{eq:affine_coupling}
\end{equation}
where $s_\theta$ and $t_\theta$ are neural networks and $\odot$ is elementwise multiplication. The block is invertible in a single pass, and its Jacobian is triangular with log-determinant $\sum_{j\in B_2} s_{\theta,j}(z_{B_1})$, linear in the dimension. Coupling blocks are stacked with the split $(B_1,B_2)$ alternated between layers so that all coordinates are eventually transformed~\cite{dinh2015nicenonlinearindependentcomponents,dinh2017densityestimationusingreal}.

\subsubsection{Overall architecture}
\label{subsec:overall_architecture}

We target probability spaces with $\mathcal{O}(10^2)$ uncertainty parameters. In this regime, a fully autoregressive architecture over all dimensions can become prohibitively expensive. At the same time, we can exploit strong a priori knowledge about the structure of a typical near-detector likelihood. Only a subset of systematics induces highly non-linear, shape-changing responses through spline-based or otherwise non-trivial reweighting, while many others act closer to linear. We therefore partition the nuisance parameters into two groups,
\begin{equation}
\boldsymbol{\eta}=(\boldsymbol{\eta}_A,\boldsymbol{\eta}_B),
\end{equation}
where $\boldsymbol{\eta}_A$ collects the non-linear parameters expected to generate complex, non-Gaussian likelihood features, and $\boldsymbol{\eta}_B$ contains the larger set of linear parameters whose effect on the likelihood is closer to linear.

We model the joint density by combining two flow models that allocate expressivity and computational cost where it matters most, using the conditional probability identity as in Eq.~\eqref{eq:results_hybrid_factorization}.

In practice, the conditioning is implemented by feeding $\boldsymbol{\eta}_B$ to the autoregressive conditioner networks, so that the parameters of the one-dimensional transformers used for $\boldsymbol{\eta}_A$ depend on $\boldsymbol{\eta}_B$. This preserves non-trivial cross-correlations while keeping the computationally expensive autoregressive structure confined to the smaller block $\boldsymbol{\eta}_A$. The implementation is based on the \texttt{normflows} package~\cite{Stimper2023}, with a custom hybrid structure adapted to the likelihood model.

Figure~\ref{fig:architecture_schematic} summarizes the corresponding implementation as a composition of three invertible stages applied to base variables $u$ drawn from a tractable distribution. We take
\begin{equation}
u \sim p_u,
\qquad
p_u(u)=\mathcal{N}(0,I),
\end{equation}
i.e.\ a product of independent standard normal Gaussians. We first apply a multivariate linear transformation
\begin{equation}
x \;=\; T^{(1)}_{\theta_1}(u) \;=\; L_{\theta_1}\,u + \mu_{\theta_1},
\label{eq:linear_flow}
\end{equation}
with $\mu_{\theta_1}\in\mathbb{R}^{D_A+D_B}$ and $L_{\theta_1}$ lower-triangular. This layer can be treated as a fixed reparametrization or trained jointly with the remaining flow parameters. We initialize it to match the post-fit Gaussian approximation: MINUIT~\cite{MINUIT} provides the best-fit point $\hat{\boldsymbol{\eta}}$ and HESSE provides the Hessian $H$ of $-\ln\mathcal{L}$ at $\hat{\boldsymbol{\eta}}$, from which $C=H^{-1}$. We then set
\begin{equation}
\mu_{\theta_1}=\hat{\boldsymbol{\eta}},
\qquad
C = L_{\theta_1}L_{\theta_1}^\top,
\end{equation}
so that the first stage reproduces the Gaussian post-fit mean and covariance. Alternatively, $\mu_{\theta_1}$ and $L_{\theta_1}$ can be initialized based on any convenient Gaussian model (e.g.\ the prior model), and optionally kept fixed.

The subsequent non-linear stages implement the hybrid factorization of Eq.~\eqref{eq:results_hybrid_factorization} by splitting $x=(x_A,x_B)$ according to the same partition as $\boldsymbol{\eta}$. The easy, high-dimensional block is modeled with a coupling-layer flow,
\begin{equation}
\boldsymbol{\eta}_B \;=\; T^{(2)}_{\theta_2}(x_B),
\end{equation}
while the difficult block is modeled with a conditional autoregressive flow,
\begin{equation}
\boldsymbol{\eta}_A \;=\; T^{(3)}_{\theta_3}(x_A;\boldsymbol{\eta}_B),
\end{equation}
where the dependence on $\boldsymbol{\eta}_B$ encodes the conditioning.

The model can be used in both directions required for our application. For sampling, we draw $u\sim p_u$, compute $x=T^{(1)}_{\theta_1}(u)$, then obtain $\boldsymbol{\eta}_B$ and finally $\boldsymbol{\eta}_A$ conditionally, returning $\boldsymbol{\eta}=(\boldsymbol{\eta}_A,\boldsymbol{\eta}_B)$. For density evaluation, Eq.~\eqref{eq:results_hybrid_factorization} yields the decomposition
\begin{equation}
\log q_{\rm NF}(\boldsymbol{\eta}_A,\boldsymbol{\eta}_B)
=
\log q_{\rm AR}(\boldsymbol{\eta}_A\mid \boldsymbol{\eta}_B)
+
\log q_{\rm CL}(\boldsymbol{\eta}_B),
\end{equation}
so the two contributions are computed with the architectures best suited to their respective blocks. Since in realistic target applications we expect $\dim(\boldsymbol{\eta}_B)\gg \dim(\boldsymbol{\eta}_A)$, this hybrid design retains the expressivity needed to capture strongly non-Gaussian directions, while keeping both sampling and likelihood evaluation tractable in high dimensions.

\subsection{Training}
\subsubsection{Loss function}

Normalizing flows are trained by adjusting the parameters $\theta$ of a tractable surrogate density $q_\theta$ so that it matches a target distribution $p$. A standard objective is based on the Kullback--Leibler divergence (KL), which quantifies the discrepancy between two probability densities. Since the KL divergence is asymmetric, minimizing it can lead to qualitatively different behaviors depending on the direction.

Minimizing the \emph{forward} KL divergence,
\begin{equation}
D_{\mathrm{KL}}(p\|q_\theta)
= \mathbb{E}_{x\sim p}\!\left[\log\frac{p(x)}{q_\theta(x)}\right],
\end{equation}
typically yields \emph{mass-covering} (or mean-seeking) solutions: the learned density $q_\theta$ is encouraged to assign probability wherever $p$ has support, which can result in overly diffuse tails. By contrast, minimizing the \emph{reverse} KL divergence,
\begin{equation}
D_{\mathrm{KL}}(q_\theta\|p)
= \mathbb{E}_{x\sim q_\theta}\!\left[\log\frac{q_\theta(x)}{p(x)}\right],
\end{equation}
often produces \emph{mode-seeking} behavior: $q_\theta$ may concentrate on a subset of dominant modes while underestimating (or ignoring) other relevant regions of the target.

To mitigate these complementary pathologies, we consider a symmetric discrepancy defined as the average of the two KL divergences,
\begin{equation}
D_{\mathrm{s}}(p,q_\theta)
\;=\;\frac{1}{2}\Big(D_{\mathrm{KL}}(p\|q_\theta)+D_{\mathrm{KL}}(q_\theta\|p)\Big).
\label{eq:symmetric_kl_def}
\end{equation}

A direct evaluation of Eq.~\eqref{eq:symmetric_kl_def} is impractical in our setting. The forward term requires samples from the target $p$, which are expensive to obtain, while the reverse term requires sampling from $q_\theta$, which can be highly unstable early in training. To obtain a stable estimator, we rely on importance sampling with a proposal density $q$ that is easy to sample from and sufficiently covers the relevant support of the target~\cite{DBLP:journals/corr/abs-1808-03856}. The specific choice and adaptation of $q$ are discussed in Section~\ref{Iterative training}.

Let $x_i\sim q$ for $i=1,\dots,N$. By importance sampling, the symmetric KL in Eq.~\eqref{eq:symmetric_kl_def} can be approximated using the Monte Carlo estimator:
\begin{equation}
D_{\mathrm{s}}(p,q_\theta)
\;\approx\;
\frac{1}{2N}\sum_{i=1}^N
\left[
\frac{p(x_i)}{q(x_i)}\log\frac{p(x_i)}{q_\theta(x_i)}
+\frac{q_\theta(x_i)}{q(x_i)}\log\frac{q_\theta(x_i)}{p(x_i)}
\right],
\label{eq:symmetric_kl_is_estimator}
\end{equation}
where the sampling is performed from the proposal distribution $q(x_i)$, and $p(x_i)/q(x_i)$ and $q_\theta(x_i)/q(x_i)$  are the importance weights.

This estimator is robust provided that $q$ adequately covers the regions where both $p$ and $q_\theta$ have significant mass. In practice, such coverage prevents the flow from collapsing onto narrow subsets of the manifold and reduces the risk of missing important modes. Nevertheless, the variance of Eq.~\eqref{eq:symmetric_kl_is_estimator} can still be large if $q$ is poorly matched to the target, which motivates the iterative construction of increasingly accurate proposals described in Section~\ref{Iterative training}.

A final subtlety concerns the fact that the target likelihood introduced in Section~\ref{app:likelihood_construction} is available only up to a multiplicative constant. Denoting by $\tilde p(x)$ the unnormalized target, we have
\begin{equation}
p(x)=\frac{\tilde p(x)}{Z},
\qquad
Z=\int \tilde p(x)\,dx.
\end{equation}
We estimate the normalizing constant $Z$ via importance sampling under the same proposal $q$,
\begin{equation}
\hat Z
=\frac{1}{N}\sum_{i=1}^N \frac{\tilde p(x_i)}{q(x_i)}.
\label{eq:Z_is_estimate}
\end{equation}
$\hat Z$ enters the loss function as a normalizing factor for the target distribution $p(x_i)$.

\subsubsection{Iterative training}
\label{Iterative training}

Considering the importance-sampling-based definition of the loss function (Eq.~\eqref{eq:symmetric_kl_is_estimator}), and the large dimensionality of the space, careful choice of the proposal distribution is required for efficient training. The target density can be far from analytically simple, and using a fixed proposal (e.g.\ the prior or posterior Gaussian) can lead to very large-variance importance weights~\cite{importance_sampling}. Sampling directly from the current flow $q_\theta$ at every gradient step is also undesirable: it would require evaluating the target likelihood online for each newly generated sample, effectively turning training into an MCMC-like procedure with a prohibitive wall time as dimensionality increases.

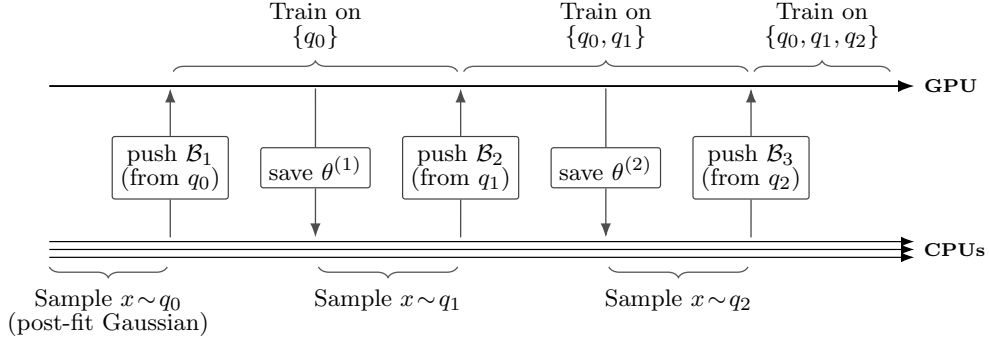
\begin{figure}[t]
\centering
\begin{tikzpicture}[
scale=0.8,
  >=Latex,
  axisline/.style={line width=0.7pt, ->, >=Latex},
  cpulane/.style ={line width=0.45pt, ->, >=Latex},
  evtbox/.style  ={draw=black!75, line width=0.45pt, rounded corners=1pt,
                   align=center, inner sep=3pt, fill=white, font=\scriptsize},
  ar/.style      ={->, line width=0.5pt, >=Latex, draw=black!70},
  noar/.style    ={line width=0.5pt, draw=black!70},
  hdr/.style     ={font=\footnotesize\bfseries},
  plabel/.style  ={align=center, font=\scriptsize},
  brG/.style     ={decorate, decoration={brace, amplitude=4pt, raise=2pt},
                   line width=0.5pt, draw=black!65},
  brC/.style     ={decorate, decoration={brace, amplitude=4pt, mirror, raise=2pt},
                   line width=0.5pt, draw=black!65},
]

\def\yG{2.7}\def\yC{0}\def\yE{1.35}
\def\xL{-0.3}\def\xT{14.0}
\def\xa{1.7}\def\xb{4.1}\def\xc{6.5}\def\xd{8.9}\def\xe{11.3}

\draw[axisline] (\xL,\yG) -- (\xT,\yG) node[right, hdr]{GPU};
\foreach \dy in {-0.13,0,0.13}
  \draw[cpulane] (\xL,\yC+\dy) -- (\xT,\yC+\dy);
\node[right, hdr] at (\xT,\yC) {CPUs};

\draw[brG] (\xa+0.05,\yG+0.15) -- (\xc-0.05,\yG+0.15)
  node[midway, above=7pt, plabel]{Train on\\$\{q_0\}$};
\draw[brG] (\xc+0.05,\yG+0.15) -- (\xe-0.05,\yG+0.15)
  node[midway, above=7pt, plabel]{Train on\\$\{q_0,q_1\}$};
\draw[brG] (\xe+0.05,\yG+0.15) -- (\xT-0.4,\yG+0.15)
  node[midway, above=7pt, plabel]{Train on\\$\{q_0,q_1,q_2\}$};

\draw[brC] (\xL,\yC-0.2)       -- (\xa-0.05,\yC-0.2)
  node[midway, below=7pt, plabel]{Sample $x\!\sim\!q_0$\\(post-fit Gaussian)};
\draw[brC] (\xb+0.05,\yC-0.2)  -- (\xc-0.05,\yC-0.2)
  node[midway, below=7pt, plabel]{Sample $x\!\sim\!q_1$};
\draw[brC] (\xd+0.05,\yC-0.2)  -- (\xe-0.05,\yC-0.2)
  node[midway, below=7pt, plabel]{Sample $x\!\sim\!q_2$};

\node[evtbox] (E1) at (\xa,\yE) {push $\mathcal{B}_1$\\(from $q_0$)};
\draw[noar] (\xa,\yC+0.2) -- (E1.south);
\draw[ar]   (E1.north) -- (\xa,\yG-0.04);

\node[evtbox] (E3) at (\xc,\yE) {push $\mathcal{B}_2$\\(from $q_1$)};
\draw[noar] (\xc,\yC+0.2) -- (E3.south);
\draw[ar]   (E3.north) -- (\xc,\yG-0.04);

\node[evtbox] (E5) at (\xe,\yE) {push $\mathcal{B}_3$\\(from $q_2$)};
\draw[noar] (\xe,\yC+0.2) -- (E5.south);
\draw[ar]   (E5.north) -- (\xe,\yG-0.04);

\node[evtbox] (E2) at (\xb,\yE) {save $\theta^{(1)}$};
\draw[noar] (\xb,\yG-0.04) -- (E2.north);
\draw[ar]   (E2.south) -- (\xb,\yC+0.2);

\node[evtbox] (E4) at (\xd,\yE) {save $\theta^{(2)}$};
\draw[noar] (\xd,\yG-0.04) -- (E4.north);
\draw[ar]   (E4.south) -- (\xd,\yC+0.2);

\end{tikzpicture}
\caption{Timeline of the asynchronous training loop.
The GPU (top) trains continuously on the FIFO buffer~$\mathcal{D}$ and
periodically writes a flow checkpoint $\theta^{(k)}$. CPU workers (bottom)
run in parallel: each samples $x\!\sim\!q_k$ from the most recent checkpoint,
evaluates the unnormalized target $\tilde p$, and pushes a batch
$\mathcal{B}_j=\{x_i,\tilde p(x_i),q_j(x_i)\}_{i=1}^{N_j}$ into~$\mathcal{D}$,
where $q_j$ is the proposal that produced batch~$j$. We use the shorthand
$q_k\!\equiv\!q_{\theta^{(k)}}$ for $k\!\geq\!1$; $q_0$ is the initial
post-fit Gaussian. Once $\mathcal{D}$ contains $K$ batches, each
new push evicts the oldest.}
\label{fig:training_schematic}
\end{figure}

We therefore adopt an iterative, asynchronous scheme in which proposal samples and target evaluations are generated on CPU workers from a lagged flow checkpoint, while GPU training proceeds continuously on the most recent aggregated dataset. Concretely, the procedure is as follows.

We maintain a dataset $\mathcal{D}$ consisting of a queue of the last $K$ batches,
\begin{equation}
\mathcal{D} = \bigcup_{j=1}^{K} \mathcal{B}_j,
\qquad
\mathcal{B}_j = \{x^{(j)}_i,\ \tilde p(x^{(j)}_i),\ q(x^{(j)}_i)\}_{i=1}^{N_j},
\end{equation}
where each batch $\mathcal{B}_j$ contains proposal samples $x^{(j)}_i$ together with their unnormalized target evaluations $\tilde p(x^{(j)}_i)$ and proposal densities $q(x^{(j)}_i)$ (needed for importance weighting). The queue is updated in FIFO fashion: when a new batch $\mathcal{B}_{K+1}$ is produced, the oldest batch is discarded so that only the last $K$ batches are kept. This controls memory and prevents the training distribution from being dominated by outdated proposals.

Batches are generated asynchronously. At regular intervals, we save a checkpoint of the current flow parameters, denoted $\theta^{\mathrm{ckpt}}$. CPU workers load this checkpoint, sample from the corresponding flow, and use that distribution as the proposal,
\begin{equation}
q(x) \equiv q_{\theta^{\mathrm{ckpt}}}(x).
\end{equation}
For each proposed point $x_i\sim q_{\theta^{\mathrm{ckpt}}}$, the workers evaluate the expensive likelihood to obtain $\tilde p(x_i)$. The resulting batch is then pushed into the FIFO queue and becomes available to the trainer.

The GPU trainer continuously optimizes $\theta$ using the current dataset $\mathcal{D}$. Since $\mathcal{D}$ is a union of batches that may have been generated by different checkpoints (hence different proposals), the loss is computed as a weighted average over batches:
\begin{equation}
\widehat{D}_{\mathrm{s}}(p,q_\theta;\mathcal{D})
=
\sum_{j=1}^{K} \frac{N_j}{\sum_{\ell=1}^{K} N_\ell}\;
\widehat{D}_{\mathrm{s}}^{(j)}(p,q_\theta),
\label{eq:batch_weighted_loss}
\end{equation}
with per-batch estimators
\begin{equation}
\widehat{D}_{\mathrm{s}}^{(j)}(p,q_\theta)
=
\frac{1}{2N_j}\sum_{i=1}^{N_j}
\left[
\frac{p(x^{(j)}_i)}{q_j(x^{(j)}_i)}\log\frac{p(x^{(j)}_i)}{q_\theta(x^{(j)}_i)}
+
\frac{q_\theta(x^{(j)}_i)}{q_j(x^{(j)}_i)}\log\frac{q_\theta(x^{(j)}_i)}{p(x^{(j)}_i)}
\right],
\label{eq:per_batch_symmetric_kl}
\end{equation}
where $q_j$ denotes the proposal that generated batch $\mathcal{B}_j$ (i.e.\ the checkpoint flow used by the workers when producing that batch). In particular, when the proposal is taken to be a checkpointed flow, we store $q_j(x^{(j)}_i)=q_{\theta^{(j)}}(x^{(j)}_i)$ alongside each sample to avoid re-evaluating old proposals.

This iterative procedure makes the proposal progressively closer to the target as training improves, which reduces the variance of the importance weights and stabilizes optimization. The overall workflow, including the checkpointing, parallel sampling, FIFO dataset update, and continuous optimization, is summarized in Figure~\ref{fig:training_schematic}. Table~\ref{tab:hyperparams} summarizes the tunable parameters of the iterative training, alongside the other normalizing flows model hyperparameters.

\begin{table}[t]
\centering
\begin{tabular}{lll}
\hline
 & Parameter & Value \\
\hline
\multirow{6}{*}{Architecture}
 & Coupling blocks ($q_{\rm CL}$)              & 12 \\
 & Autoregressive layers ($q_{\rm AR}$)        & 12 \\
 & Spline bins                                  & 9 \\
 & Spline tail bound $B$                        & 6 \\
 & Coupling conditioner network                 & 1 layer $\times$ 256 units \\
 & Autoregressive conditioner network           & 2 layers $\times$ 512 units \\
\hline
\multirow{5}{*}{Training}
 & Optimizer                                    & Adam (AMSGrad) \\
 & Learning rate (constant)                                & $1\times10^{-5}$ \\
 & Sampling batch size $N_j$                     & $10\,000$ \\
 & FIFO buffer length                           & 50 batches \\
 & Parallel CPU samplers                         & 10 \\
\hline
\end{tabular}
\caption{Hyperparameters of the hybrid normalizing-flow architecture and of the training procedure. The coupling and autoregressive blocks refer to the two factors of Eq.~\eqref{eq:results_hybrid_factorization}; the spline bins and tail bound $B$ specify the rational-quadratic transform of Eq.~\eqref{eq:ar_rqnsf}, which acts as the identity outside the interval $[-B,B]$. Each conditioner network is given as (hidden layers $\times$ units per layer). The FIFO buffer length is the maximum number $K$ of batches of size $N_j$ retained in the training dataset $\mathcal{D}$, and the parallel CPU samplers are the workers that generate proposal points from the asynchronous checkpoint (Section~\ref{Iterative training}).}
\label{tab:hyperparams}
\end{table}

In the present analysis, the initial training dataset is generated by sampling the systematic parameters $\boldsymbol{\eta}$ from $g(\boldsymbol{\eta};\boldsymbol{\eta}_{\mathrm{best\text{-}fit}},C)$. Since these initial samples are drawn independently from the post-fit Gaussian distribution, their generation and likelihood evaluation are embarrassingly parallelizable and can therefore be performed efficiently. Training starts from an initial dataset of $2.5$ million vectors of systematic parameters, together with their corresponding likelihood evaluations. In this procedure, $10$ CPU workers continuously generate new points from the NF proposal and compute their corresponding likelihood values. After $64\,000$ epochs, corresponding to approximately $26$ hours of training and an additional $1.5$ million training points, the procedure is stopped and the final NF approximation of the likelihood is obtained.

\bmhead{Data availability}
Data supporting the findings of this study are openly
available in Zenodo at \href{https://doi.org/10.5281/zenodo.21104524}{https://doi.org/10.5281/zenodo.21104524}.

\bmhead{Code availability}
The code used in this study is openly available on GitHub. The normalizing-flow surrogate, including the hybrid architecture and training loop, is at \href{https://github.com/GunFlows/gunflows}{https://github.com/GunFlows/gunflows}, and the benchmark near-detector likelihood and fit are at \href{https://github.com/GunFlows/ToyNDFit}{https://github.com/GunFlows/ToyNDFit}. The flow implementation builds on the open-source \texttt{normflows} package~\cite{Stimper2023}. Likelihood evaluations are performed with the GUNDAM framework~\cite{gundam}.

\backmatter

\bmhead{Acknowledgements}
We thank S.~Samani, K.~Vedantha, S.~Bordoni, and L.~Amziane for their valuable contributions and support. We thank S.~Dolan and C.~Wilkinson for providing datasets of simulated neutrino-nucleus interactions used for the demonstrator near-detector fit.

\bmhead{Funding}

This work was partially supported by the Swiss National Science Foundation (SNSF) under grants No. 200021E\_213196 and No. 200020\_204609. 

\bmhead{Author contributions}
M.E.B. and L.G. conceived the use of normalizing flows as high-dimensional likelihood surrogates, developed the normalizing-flow surrogate model and its training, as well as the benchmark near-detector fit replica, and carried out the analysis. A.B. contributed to the statistical methodology and software implementation, including the interface with the GUNDAM software. F.S. supervised every aspect of the project. All authors discussed the results and wrote, reviewed and approved the manuscript. 

\bmhead{Competing interests}
The authors declare no competing interests.

\bibliography{sn-bibliography}% common bib file

\end{document}